\documentclass[useAMS]{mn2e}
\usepackage{graphicx}
\usepackage{longtable}

\title[AGB Stars in the Fornax Dwarf Spheroidal Galaxy]{AGB Stars 
in the Fornax Dwarf Spheroidal Galaxy}
\author[P. A. Whitelock et al.]{Patricia A. Whitelock$^{1,2,3}$, 
John W. Menzies$^1$, Michael W. Feast$^2$, Noriyuki \newauthor
Matsunaga$^4$, Toshihiko Tanab\'e$^5$ and Yoshifusa Ita$^6$\\
      $^1$ South African Astronomical Observatory, P.O.Box 9, 7935
           Observatory, South Africa.\\
      $^2$ National Astrophysics and Space Science Programme, 
	   Astronomy Department, University of Cape Town, 7701 Rondebosch,\\
           South Africa.\\
      $^3$ Department of Mathematics and Applied Mathematics,
           University of Cape Town, 7701 Rondebosch, South Africa.\\
      $^4$ Department of Astronomy, Kyoto University, Kitashirakawa
           Oiwake-cho, Sakyo-ku, Kyoto, Kyoto 606-8502, Japan.\\
      $^5$ Institute of Astronomy, School of Science, The University of Tokyo, 
           Mitaka, Tokyo 181-0015, Japan.\\
      $^6$ Institute of Space and Astronautical Science, Japan Aerospace
           Exploration Agency, 3-1-1 Yoshinodai, Sagamihara,\\ Kanagawa
           229-8510, Japan. }
\begin{document}
\maketitle
\begin{abstract} We report on a multi-epoch study of the Fornax
dwarf spheroidal galaxy, made with the Infrared Survey Facility, over an
area of about $42' \times 42'$. The colour-magnitude diagram shows a broad
well-populated giant branch with a tip that slopes down-wards from red to
blue, as might be expected given Fornax's known range of age and
metallicity. The extensive AGB includes seven Mira variables and ten
periodic semi-regular variables. Five of the seven Miras are known to be
carbon rich. Their pulsation periods range from 215 to 470 days, indicating
a range of initial masses. Three of the Fornax Miras are redder than typical
LMC Miras of similar period, probably indicating particularly heavy
mass-loss rates. Many, but not all, of the characteristics of the AGB are
reproduced by isochrones from Marigo et al. for a 2 Gyr population with a
metallicity of Z=0.0025.

An application of the Mira period-luminosity relation to these stars yields
a distance modulus for Fornax of 20.69 $\pm 0.04$ (internal), $\pm 0.08$
(total) (on a scale that puts the LMC at 18.39 mag) in good agreement with
other determinations. Various estimates of the distance to Fornax are
reviewed.
 
\end{abstract}

\section{Introduction}

The Fornax dwarf spheroidal galaxy ($\l=237^{\rm o}.24$, $b=-65^{\rm o}.66$)
is one of the most populous dwarf spheroidal companions to the Milky Way,
second only to the disrupting Sagittarius dwarf. It shows evidence for a
very extended history of star formation that ceased only a few million years
ago (e.g. Gallart et al. 2005; Coleman \& de Jong 2008) and a significant
range of metallicity (e.g. Battaglia et al. 2006). It has its own system of
five metal deficient globular clusters (Buonanno et al. 1999; Letarte et al.
2006), an extended asymptotic giant branch (AGB) with numerous carbon stars
(e.g. Westerlund, Edvardsson \& Lundgren 1987; Lundgren 1990) and evidence
of substructure, possibly a consequence of merging with another galaxy
(Coleman et al. 2004; Olszewski et al. 2006). The young stars are more
concentrated towards the centre of the galaxy (Coleman \& de Jong 2008) and
there is an anti-correlation between the iron content and the velocity
dispersion (Battaglia et al. 2006), suggestive of an age metallicity
relation.

This paper is one of a series aimed at finding and characterizing luminous
AGB variables within Local Group galaxies; it follows similar work on Leo I
and Phoenix (Menzies et al. 2002, 2008). Here we report on multi-epoch
$JHK_S$ photometry which enables us to identify the AGB variables within
Fornax. The large-amplitude variables, generally known as Miras are of
particular interest, first, because they tell us about the intermediate age
population of which they are the most luminous representatives, secondly
because they provide an independent distance calibration and thirdly because
they will be major contributors to the processed material currently
entering the interstellar medium of Fornax and are therefore an important
source of enrichment.

Estimates for the interstellar extinction towards Fornax are mostly in the
range $0.03<E(B-V)<0.1$ mag (e.g. Greco et al. 2007 and references therein).
Here we use a value near the middle of this range, $E(B-V)=0.07$ mag, which
amounts to $A_K<0.02$ and is therefore almost negligible in our work.

\begin{figure} 
\includegraphics[width=8.5cm]{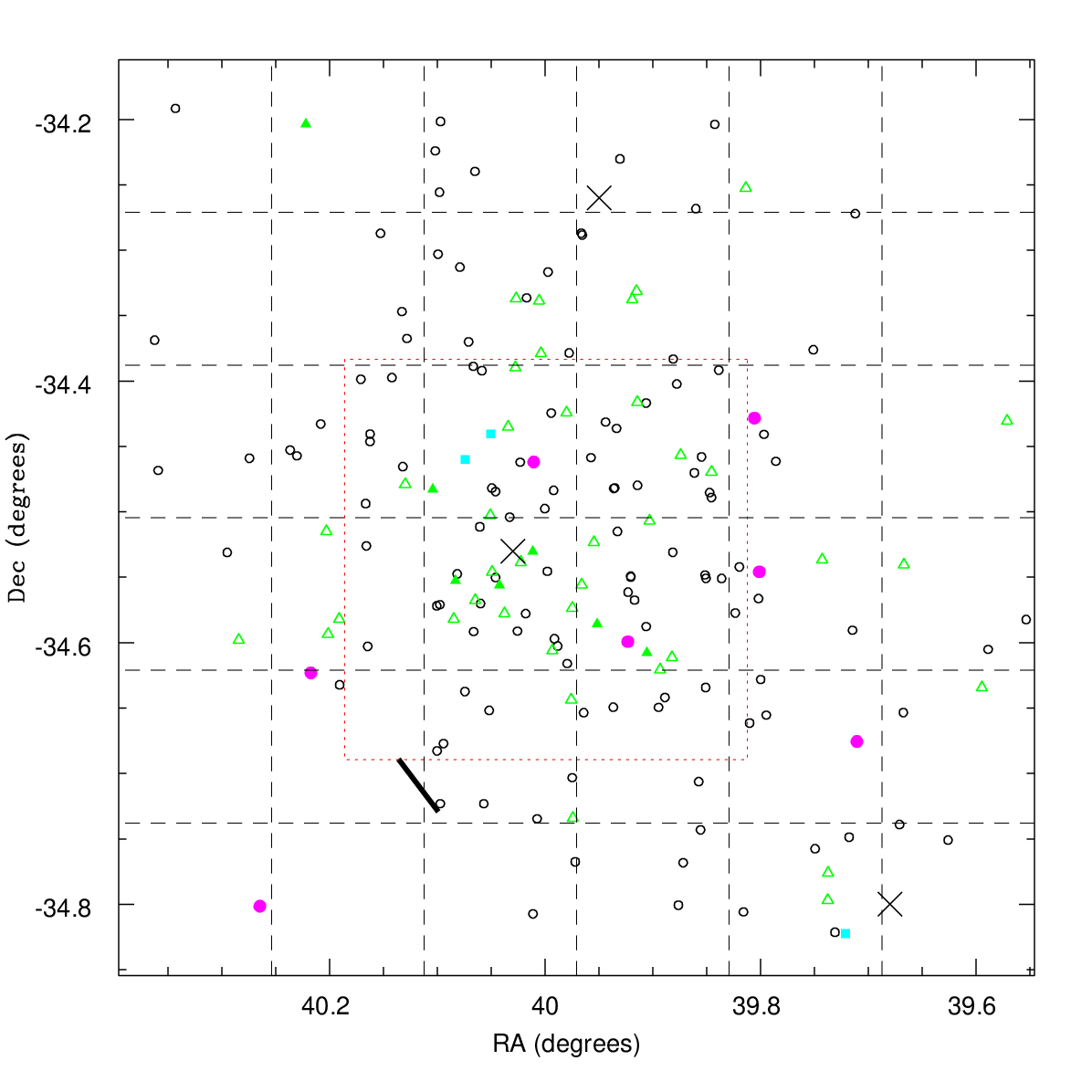}
 \caption{The fields covered by our survey and the
distribution of the AGB variables and AGB candidates. Periodic variable
stars are indicated as solid symbols: circles (Miras), squares (SRs with
long term trends) and triangles (SRs). Variables with undetermined periods
are shown as open triangles. Globular clusters (Hodge 1961) are
shown as large crosses and the overdensity noted by Coleman et al. (2004) is
shown as a bar.  The centroid of Fornax, from van den Bergh (2000), is at
the centre of the image (RA: $\rm 02^h\ 39^m\ 53^s$, Dec: $\rm -34^o\ 30'\
16''$ equinox 2000). The dashed lines mark the boundaries of our fields and
the dotted line shows the region covered by the shallow survey of
Gullieuszik et al. (2008).}
 \label{fig_pos}
\end{figure}

\section{Observations}
 Observations were made with the SIRIUS camera on the Japanese-SAAO Infrared
Survey Facility (IRSF) at Sutherland. The camera produces simultaneous $J$,
$H$ and $K_S$ images covering what is typically a $7.2 \times 7.2$ arcmin square
field (after dithering) with a scale of 0.45 arcsec/pixel. To cover most of
the Fornax galaxy, a $6 \times 6$ grid was used, with successive image
centres displaced by 7 arcmin in RA along a row and successive rows
displaced by 7 arcmin in Dec (see Fig.~\ref{fig_pos}).

Since the aim was to find long-period variables, observations were made at
about 15 epochs spread over 3 years in the central $4\times 4$ 
grid; poor weather and other constraints reduced this number to about 11 in
the outer ring of 20 fields. In each field, 10 dithered images were combined
after flat-fielding and dark and sky subtraction. Typical exposures were of
either 10 or 20\,s, depending on the seeing and on the brightness of the sky
in the
$K_S$ band. Photometry was performed using DoPHOT (Schechter et al. 1993)
in `fixed-position' mode, using the best-seeing $H$-band image in each field
as templates. Aladin was used to correct the coordinate system on each
template and RA and Dec were determined for each measured star. This allowed
a cross-correlation to be made with the 2MASS catalogue (Cutri et al. 2003),
and photometric zero points were determined by comparison of our photometry
with that of 2MASS. In each field, stars in common with the 2MASS catalogue
with photometric quality A in each colour were identified and the IRSF zero
point was adjusted to match that of 2MASS. The number of common stars per
field varies from over 20 in the middle four fields to as few as four in
field 26; only 5 fields have fewer than 7 stars in common with 2MASS. The
mean standard deviation over all fields of the differences between IRSF and
2MASS are 0.06 mag in $J$ and $H$ and 0.08 mag in $K_S$. No account was
taken of possible colour transformations, such as in Kato et al. (2007).
Those transformations were derived using highly reddened objects to define
the red end and it is not obvious that the same transformations will apply
to carbon stars.

For a given field, 10 dithered frames were median averaged to produce an
image for measurement.  Stars near the edge of a frame tend to have lower
photometric precision since they might not have appeared on all 10
dithered frames. Variables appearing in the overlap regions between
fields may have been missed because of this effect, though this is
unlikely for large amplitude very red variables.

\section{Colour-Magnitude and Colour-Colour Diagrams}

\begin{figure}
\includegraphics[width=8.5cm]{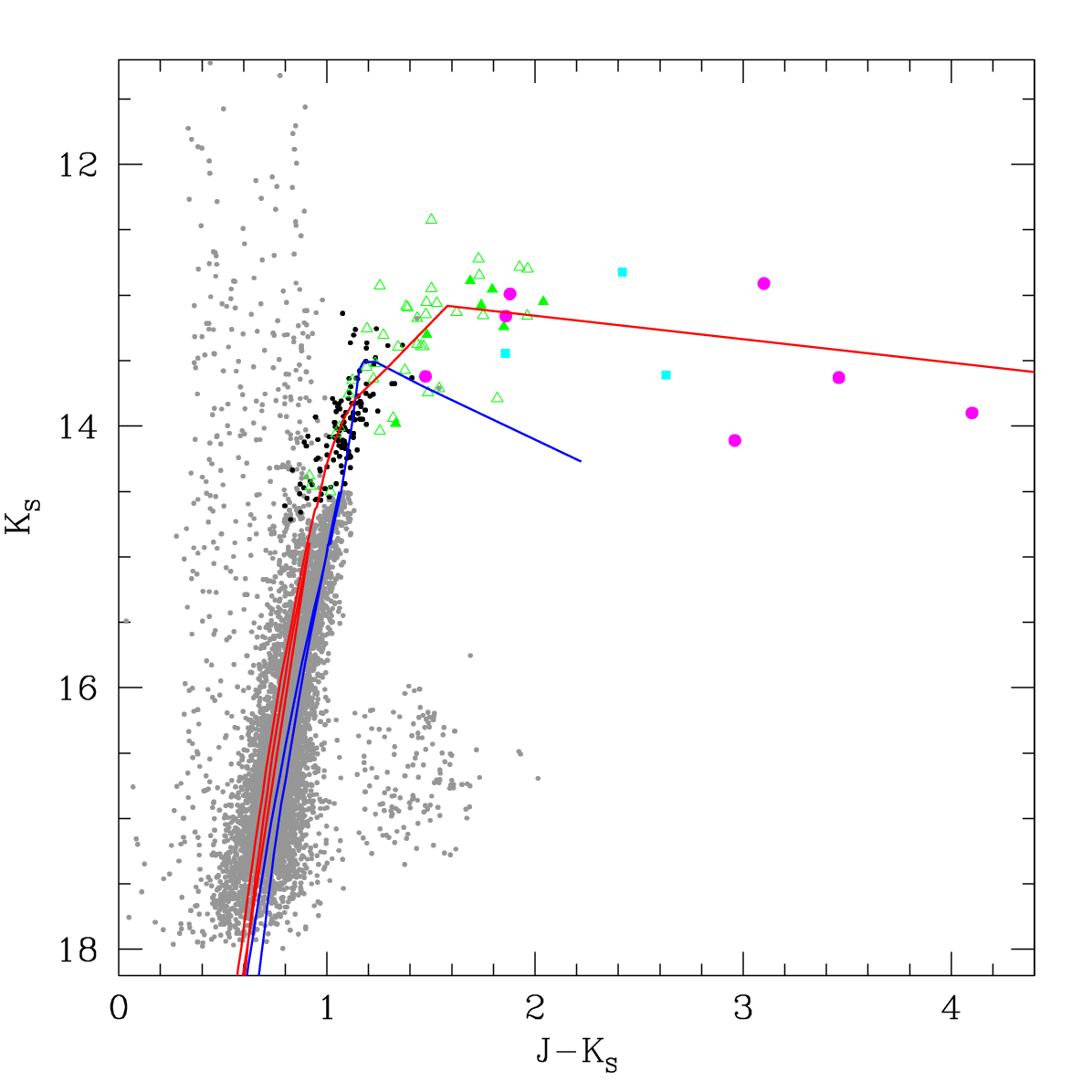}
\caption{Colour-magnitude diagram for Fornax; large circles are Mira
variables, squares are SRs with long term trends, triangles are low
amplitude, SR or Irr, variables, while small black circles are other AGB
stars. The curves are isochrones from Marigo et al. (2008) for populations
with ages of 2 and 10\,Gyr (details in text). }
\label{fig_cm}
\end{figure}

\begin{figure}
\includegraphics[width=8.5cm]{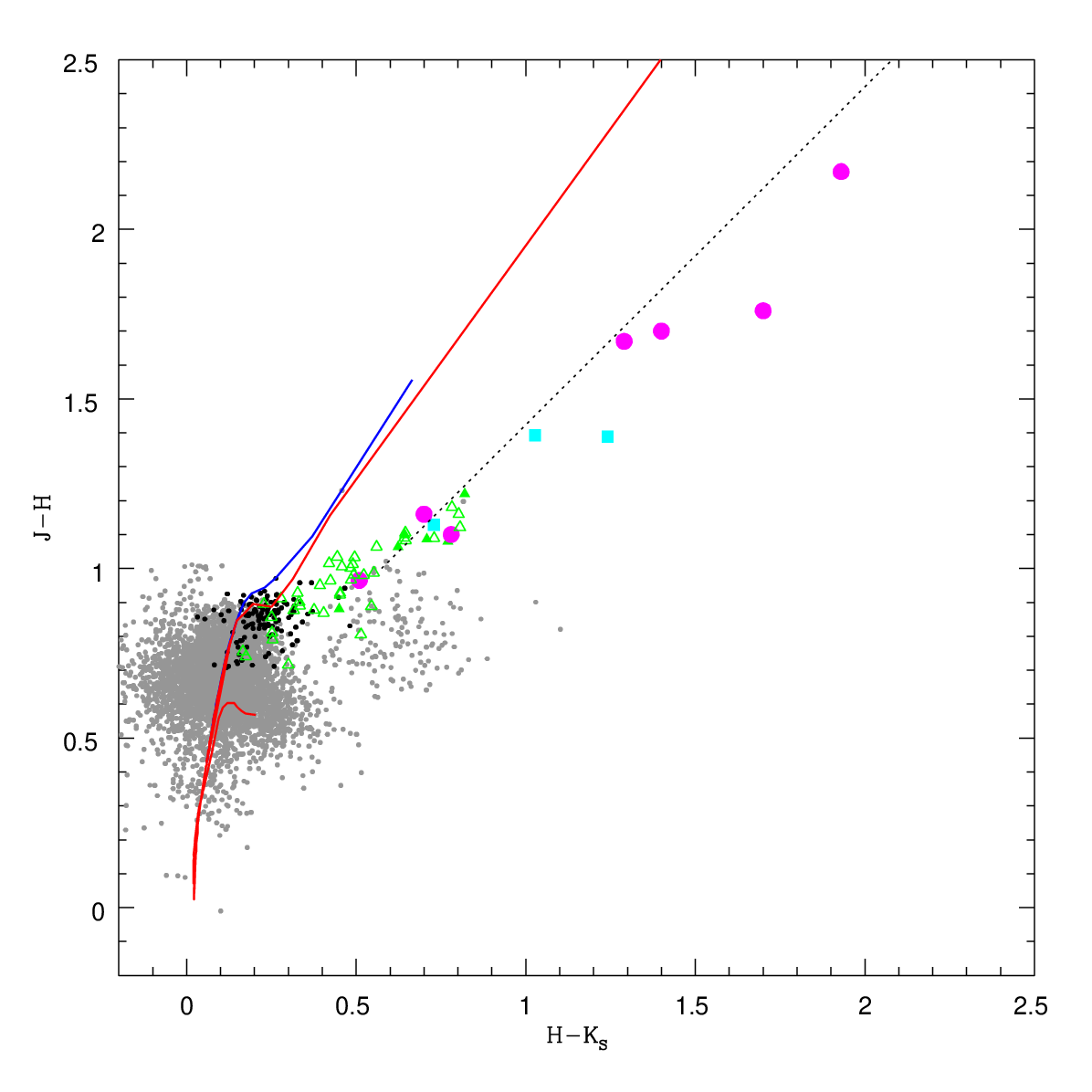}
\caption{Two colour diagram for Fornax. Symbols as in Fig.~\ref{fig_cm}.
The dotted line represents the relation for Galactic carbon Miras
(equation 2 from Whitelock et al. (2006)) approximately converted onto the
2MASS system (2MASS: $ (H-K)_0=-0.428+1.003(J-H)_0$). The other two lines
are the same models as shown in Fig.~\ref{fig_cm}. }
\label{fig_jhhk1}
\end{figure}

Fig.~\ref{fig_cm} shows the $K_S-(J-K_S)$ diagram and
Fig.~\ref{fig_jhhk1} the $(J-H)-(H-K_S)$ diagram for stars selected as 
follows: for bright stars, i.e. those with $J<16$, $H<15.5$ or $K_S<15$ mag,
standard deviations $\sigma < 0.11$ mag; for fainter stars the
limit depends on the magnitude: for   
$16 <J <19$ it is $\sigma<0.1J-1.5 $, and for $15.5 <H <18.5$ it is $\sigma<0.1H-1.45 $ and 
for $15 <K_S <18$ it is $\sigma<0.15K-2.15 $. Mean magnitudes from all of
our observations are used in all of these plots.
Selecting the stars in this way is intended to reject any poor quality
photometry, but it also eliminates large amplitudes variables. The latter are
extracted separately, by examining the light curves of the bright stars 
with large scatter, and are shown in the figures as different symbols.

In Fig.~\ref{fig_cm} foreground contamination is seen as a broad band
with $0.2<J-K_S<1.0$ over the whole magnitude range with a few bluer points 
appearing at fainter magnitudes.

Unresolved galaxies provide background contamination and we can compare with
the galaxies discussed by Morris et al. (2007), who list 2MASS magnitudes
for 228 confirmed galaxies with $cz > 900\, \rm km\,s^{-1}$ in the central 2.9
square degrees of the Fornax cluster of galaxies. These are point sources
with velocities which indicate that they are behind the Fornax galaxy
cluster. These sources, which have mean colours of $J-H=0.78$ mag and
$H-K_S=0.61$ mag and a mean $K_S=15.12$ mag, are clearly magnitude limited. 
If the source density were the same behind the Fornax dwarf spheroidal we
would expect to see about thirty in our area of 0.4 square degrees.  In fact
in the sample illustrated in Fig.~\ref{fig_cm} there is only one object with
$K<16.0$ mag that stands clear from the dwarf spheroidal giant branch. What we
do find is over 130 sources, with similar colours to the Morris et al.
sample, in the magnitude range $16 <K_S< 17.3$ mag (magnitude limited at $J$),
i.e. about 1.5 mag fainter at $K_S$ than the Morris et al. sources.  In the
two-colour diagram, Fig.~\ref{fig_jhhk1}, these sources can be seen to the
right of the giant branch sources with the same mean colours as those of the
Morris et al. sample (at 1.5 mag further the volume covered is 8 times
larger). The most probable explanation of this is that we resolved galaxies
of the brightness that were unresolved by Morris et al. and therefore
rejected them early on in our analysis that was focused on stars. Indeed
there are 2MASS sources with colours and magnitudes similar to the Morris et
al. unresolved galaxies, which we do not find because they are resolved.

The bulk of the stars fall on the Fornax giant branch with $K_S>14.5$ mag,
while the AGB is represented by a group of stars immediately above the tip
of the red giant branch (TRGB) and by variable stars extending out to 
$J-K_S>4$ mag. The AGB stars are discussed in detail below.  

It is clear that the TRGB is sloped, being brighter at the red side
($K_S\sim14.5$ mag at $J-K_S=1.07$ mag) than at the blue side ($K_S\sim14.9$
mag at $J-K_S=0.81$ mag). See also Figs.~\ref{fig_sp} and \ref{fig_cmBW}
which show the TRGB at a larger scale. This is to be expected in a galaxy
which contains stars with a range of ages and metallicities. Battaglia et
al. (2006) find evidence for a young (few hundred Myr), an intermediate age
(2 to 8 Gyr) and an old ($>10$ Gyr) population. The same authors quote
earlier work as showing that the giant branch peaks at $\rm [Fe/H]\sim
-0.9$, but has a metal poor tail extending to $\rm [Fe/H]=-2$ and a metal
rich one extending beyond $\rm [Fe/H]=-0.4$. We anticipate that our own
results will be dominated by the large intermediate age population. Two
models from Marigo et al. (2008) are overplotted on Figs.~\ref{fig_cm} and
\ref{fig_jhhk1} providing a useful qualitative comparison. A distance modulus
of 20.69 mag is assumed (see section 8 for a detailed discussion of the
distance). The isochrone which goes through the variable stars (in
Fig.~\ref{fig_cm}) is that for a population with an age of 2 Gyr and a
metallicity of Z=0.0025 ($\rm [M/H]\sim-0.88$).  For this model the TRGB is
at $K_S=14.9$ mag and $J-K_S=0.92$. mag The other curve represents a
population with an age of 10 Gyr and the same metallicity; its TRGB is at
$K_S=14.5$ mag and $J-K_S=1.06$ mag. These two curves bracket the giant
branch observations and most of the stars are probably in this 2 to 10 Gyr
age range, as others have pointed out before (e.g. Gallart et al. 2005;
Coleman \& de Jong 2008). The failure of the models to come close to the
observations of the dust enshrouded AGB stars in Fig.~\ref{fig_jhhk1} is
presumably the result of an incorrect choice of dust properties (in
particular the composition and/or particle size/shape etc.). The models
offer alternative dust compositions, but none of them results in isochrones
that fit all aspects of the observations. As we refine our understanding of
the dust chemistry it will become easier to include the appropriate
parameters in the models.

\section{Asymptotic Giant Branch}
 We limit our discussion to those stars which are above the TRGB as we have
no systematic way of determining which of the lower luminosity objects are
actually on the AGB. The selection of variables, most of
which will be AGB stars, is described below. Other probable AGB members are
selected with $K_S<16-1.43(J-K_S)$, $K_S>19-5.6(J-K_S)$ and $J-H>0.7$ mag
(this last condition being to minimize the contribution from foreground
dwarfs). These are the stars that are shown in black in Figs.~\ref{fig_cm}
and \ref{fig_jhhk1}.
Our selection criteria will certainly miss a few real AGB stars,
particularly those in the overlap parts of our fields where the photometry
was not so good, and will probably contain a few foreground stars, but the
contamination should be small.

\begin{center}
\onecolumn
\small
\begin{longtable}{ccrccrrccccccll}
\caption[Non-variable ($\sigma J, \sigma H, \sigma K_S<0.11$) upper AGB candidates.]
{Non-variable ($\sigma J\sigma H\sigma K_S <0.11$) upper AGB candidates.}\label{agb} \\
\hline
\multicolumn{3}{c}{RA}&\multicolumn{3}{c}{Dec} &
\multicolumn{1}{c}{F} & $J$ & $H$ & $K_S$ & $J-K_S$ & 2MASS & Sp & other names \\
\multicolumn{6}{c}{(equinox 2000)}& & \multicolumn{4}{c}{(mag)}\\
\hline
\endfirsthead

\hline
\multicolumn{3}{c}{RA}&\multicolumn{3}{c}{Dec} &
\multicolumn{1}{c}{F} & $J$ & $H$ & $K_S$ & $J-K_S$ & 2MASS & Sp & other names \\
\multicolumn{6}{c}{(equinox 2000)}& & \multicolumn{4}{c}{(mag)}\\
\hline
\endhead
  \multicolumn{14}{l}{{Continued on Next Page\ldots}} \\
\endfoot

\multicolumn{15}{l}{Notes: The column following RA and Dec, headed F,
contains a running number unique to the star within our system. }\\
\multicolumn{15}{l}{The column
labeled Sp contains the spectral type where this is available; other names
are as follows: numbers preceded by S are} \\
\multicolumn{15}{l}{from Stetson et al. (1998), DK from Demers \& Kunkel (1979), 
DI from Demers \& Irwin (1987), BW from Bersier \& Wood (2002),}\\
\multicolumn{15}{l}{WEL from Westerlund et al. (1987), DDB from Demers et al. (2002),
GLM from Groenewegen et al. (2008), }\\
\multicolumn{15}{l}{`for" and ``ET'' come from Battaglia et al. (2006).}   \\ 
\hline
\endlastfoot
  2& 38 &12.8 &  --34& 34& 56.6 &  29009 & 15.54 & 14.83 & 14.71 &  0.83  &02381284-3434561 &                     \\   
  2& 38 &21.3 &  --34& 36& 18.5 &  29006 & 15.39 & 14.67 & 14.52 &  0.87  &02382132-3436180 & & BW5               \\
  2& 38 &30.2 &  --34& 45&  3.3 &  31012 & 15.34 & 14.53 & 14.25 &  1.10  &02383024-3445030 & & BW9 for08\_006               \\
  2& 38 &40.2 &  --34& 39& 12.1 &  30010 & 15.53 & 14.75 & 14.44 &  1.09  &02384018-3439121 & & ET0229                    \\
  2& 38 &41.1 &  --34& 44& 20.7 &  31013 & 15.54 & 14.82 & 14.57 &  0.97  &02384113-3444205 &                     \\
  2& 38 &50.9 &  --34& 16& 19.5 &  25005 & 15.20 & 14.32 & 14.24 &  0.96  &02385093-3416192 &                     \\
  2& 38 &51.6 &  --34& 35& 25.9 &  12006 & 14.79 & 13.91 & 13.64 &  1.15  &02385157-3435253 &S2/5& DK32 BW14 S8      \\
  2& 38 &52.3 &  --34& 44& 55.2 &  32012 & 15.22 & 14.39 & 14.22 &  1.00  &02385229-3444553 & &for08\_001                    \\
  2& 38 &55.4 &  --34& 49& 16.8 &  32003 & 14.21 & 13.34 & 13.14 &  1.08  &02385547-3449166 & & BW16              \\
  2& 38 &59.9 &  --34& 45& 26.8 &  32011 & 15.29 & 14.45 & 14.26 &  1.03  &02385991-3445268 &                     \\
  2& 39 & 0.3 &  --34& 22& 34.0 &  10009 & 15.09 & 14.18 & 14.02 &  1.06  &02390030-3422334 &                     \\
  2& 39 & 8.6 &  --34& 27& 40.8 &  11009 & 14.48 & 13.57 & 13.36 &  1.11  &02390862-3427407 &                     \\
  2& 39 &10.7 &  --34& 39& 19.4 &  13010 & 14.86 & 14.10 & 13.82 &  1.04  &02391074-3439196 & & BW21 S18 DI26              \\
  2& 39 &11.2 &  --34& 26& 26.8 &  11018 & 15.15 & 14.23 & 14.04 &  1.11  &02391126-3426268 &                     \\
  2& 39 &12.0 &  --34& 37& 41.3 &  12012 & 15.23 & 14.38 & 14.14 &  1.09  &02391201-3437414 & & BW23 S20 DK30 DI24             \\
  2& 39 &12.5 &  --34& 33& 58.8 &  12008 & 15.02 & 14.16 & 13.90 &  1.13  &02391249-3433587 & & BW24 S21 DK35 DI21             \\
  2& 39 &14.5 &  --34& 39& 41.3 &  13009 & 14.73 & 13.88 & 13.58 &  1.16  &02391445-3439416 &                     \\
  2& 39 &15.8 &  --34& 48& 20.9 &  32009 & 14.82 & 13.98 & 13.79 &  1.03  &02391582-3448206 &                     \\
  2& 39 &16.7 &  --34& 32& 31.3 &  12027 & 15.47 & 14.72 & 14.48 &  0.99  &02391676-3432313 & & BW25 DI18             \\
  2& 39 &17.6 &  --34& 34& 38.5 &  12007 & 15.09 & 14.21 & 13.95 &  1.14  &02391767-3434384 &S3/2& DK34           \\
  2& 39 &20.7 &  --34& 33&  3.2 &   3023 & 14.55 & 13.66 & 13.36 &  1.19  &02392068-3433030 & M & WEL-M2          \\
  2& 39 &21.3 &  --34& 23& 29.9 &   9008 & 15.10 & 14.26 & 14.08 &  1.01  &02392133-3423292 & & BW27 S28 ET0215             \\
  2& 39 &22.2 &  --34& 12& 13.7 &  24013 & 15.46 & 14.61 & 14.55 &  0.91  &02392221-3412133 & & ET0096              \\
  2& 39 &23.0 &  --34& 29& 20.3 &   2045 & 15.38 & 14.59 & 14.45 &  0.93  &02392297-3429198 &                     \\
  2& 39 &23.4 &  --34& 29&  7.3 &   2018 & 14.59 & 13.71 & 13.40 &  1.19  &02392337-3429069 & C & WEL-C15 BW29 S30    \\
  2& 39 &24.2 &  --34& 33&  2.7 &   3024 & 15.06 & 14.22 & 13.88 &  1.18  &02392418-3433021 & C & DK33 BW30 S31   \\
  2& 39 &24.2 &  --34& 38&  3.4 &  14014 & 14.74 & 13.90 & 13.64 &  1.11  &02392423-3438027 & C & DK29 BW31       \\
  2& 39 &24.4 &  --34& 32& 53.6 &   3026 & 15.12 & 14.30 & 14.09 &  1.04  &02392438-3432530 &                     \\
  2& 39 &25.2 &  --34& 27& 28.7 &   2022 & 14.89 & 14.04 & 13.83 &  1.06  &02392514-3427283 &                     \\
  2& 39 &25.4 &  --34& 44& 35.3 &  33004 & 15.04 & 14.16 & 14.00 &  1.04  &02392535-3444353 & M & BW32 DK24       \\
  2& 39 &25.8 &  --34& 42& 22.5 &  14010 & 14.88 & 14.04 & 13.81 &  1.07  &02392578-3442220 & & BW33 S36      \\
  2& 39 &26.5 &  --34& 16&  5.9 &   9044 & 15.36 & 14.50 & 14.47 &  0.89  &02392646-3416054 & & ET0099                    \\
  2& 39 &26.8 &  --34& 28& 12.6 &   2051 & 15.51 & 14.72 & 14.56 &  0.94  &02392675-3428121 & & BW34 S38             \\
  2& 39 &29.3 &  --34& 46&  5.7 &  33016 & 15.41 & 14.69 & 14.61 &  0.80  &02392925-3446057 & & BW35              \\
  2& 39 &30.3 &  --34& 48&  2.6 &  33003 & 14.99 & 14.27 & 14.08 &  0.91  &02393033-3448026 & M & WEL-M26         \\
  2& 39 &30.7 &  --34& 24&  8.1 &   2068 & 15.13 & 14.32 & 14.07 &  1.06  &02393065-3424078 &                     \\
  2& 39 &31.5 &  --34& 22& 59.3 &   9009 & 15.12 & 14.26 & 14.08 &  1.04  &02393145-3422586 &                     \\
  2& 39 &31.6 &  --34& 31& 51.4 &   3027 & 15.01 & 14.15 & 13.93 &  1.08  &02393155-3431509 & C & WEL-C14         \\
  2& 39 &33.3 &  --34& 38& 30.7 &  14032 & 15.21 & 14.38 & 14.15 &  1.06  &02393333-3438302 &C & DK28                     \\
  2& 39 &34.7 &  --34& 38& 58.0 &  14013 & 15.04 & 14.10 & 13.63 &  1.41  &                  & C & DK17           \\
  2& 39 &37.5 &  --34& 25&  0.3 &   2025 & 14.81 & 13.94 & 13.70 &  1.11  &02393752-3425000 &                     \\
  2& 39 &37.5 &  --34& 35& 15.6 &   3043 & 15.30 & 14.42 & 14.19 &  1.10  &02393755-3435149 &                     \\
  2& 39 &39.5 &  --34& 28& 47.0 &   2020 & 14.43 & 13.56 & 13.30 &  1.13  &02393946-3428468 &                     \\
  2& 39 &40.1 &  --34& 34&  2.9 &   3020 & 15.21 & 14.33 & 14.09 &  1.13  &02394011-3434024 &                     \\
  2& 39 &40.9 &  --34& 32& 59.5 &   3053 & 15.33 & 14.46 & 14.22 &  1.11  &                 &                     \\
  2& 39 &40.9 &  --34& 32& 57.0 &   3025 & 14.99 & 14.09 & 13.83 &  1.17  &02394094-3432569 &                     \\
  2& 39 &41.5 &  --34& 33& 41.0 &   3021 & 15.00 & 14.04 & 13.68 &  1.33  &02394154-3433404 & & BW40 S61             \\
  2& 39 &43.3 &  --34& 13& 48.7 &  24011 & 15.15 & 14.28 & 14.15 &  1.00  &02394336-3413485 &                     \\
  2& 39 &43.9 &  --34& 30& 54.1 &   3029 & 15.24 & 14.41 & 14.16 &  1.07  &02394386-3430534 & & BW41              \\
  2& 39 &44.1 &  --34& 26& 10.3 &   2024 & 14.92 & 14.02 & 13.77 &  1.15  &02394408-3426101 & & BW42 S63 ET0076              \\
  2& 39 &44.5 &  --34& 28& 54.2 &   2047 & 15.37 & 14.52 & 14.30 &  1.07  &                 &                     \\
  2& 39 &44.7 &  --34& 28& 55.5 &   2019 & 14.99 & 14.16 & 13.90 &  1.09  &02394470-3428552 &                     \\
  2& 39 &44.8 &  --34& 38& 57.7 &  14012 & 14.78 & 13.90 & 13.63 &  1.14  &02394486-3438572 & M & DK27            \\
  2& 39 &46.5 &  --34& 25& 52.9 &   2061 & 15.05 & 14.31 & 14.15 &  0.90  &02394652-3425527 &                     \\
  2& 39 &49.8 &  --34& 27& 30.6 &   2056 & 15.21 & 14.33 & 14.13 &  1.08  &02394978-3427304 &                     \\
  2& 39 &51.4 &  --34& 39& 12.9 &  14031 & 15.17 & 14.34 & 14.11 &  1.05  &02395144-3439126 & & BW45 S72 DI25             \\
  2& 39 &51.8 &  --34& 17& 18.0 &   9013 & 14.93 & 14.13 & 13.89 &  1.04  &02395179-3417174  &                    \\
  2& 39 &52.0 &  --34& 17& 13.0 &   9014 & 15.29 & 14.43 & 14.23 &  1.06  &02395200-3417123  & & ET0102                   \\
  2& 39 &53.3 &  --34& 46&  2.6 &  34013 & 15.18 & 14.30 & 13.99 &  1.19  &02395331-3446026 & C &  DK10 BW47 S76       \\
  2& 39 &54.0 &  --34& 42& 11.5 &  15017 & 15.01 & 14.15 & 13.97 &  1.04  &02395400-3442111  &                    \\
  2& 39 &54.6 &  --34& 22& 42.8 &   2028 & 14.99 & 14.16 & 13.68 &  1.31  &02395462-3422426  & C & DK43 BW48 S80     \\
  2& 39 &55.1 &  --34& 36& 57.6 &   4038 & 15.17 & 14.46 & 14.34 &  0.83  &02395512-3436571  &                    \\
  2& 39 &57.2 &  --34& 36&  9.1 &   4014 & 14.89 & 14.00 & 13.79 &  1.10  &02395721-3436085  &                    \\
  2& 39 &58.0 &  --34& 35& 49.1 &   4015 & 15.20 & 14.31 & 14.12 &  1.08  &02395749-3435485  &                    \\
  2& 39 &58.1 &  --34& 29&  0.9 &   1015 & 15.02 & 14.29 & 14.12 &  0.89  &02395807-3429007 & M & WEL-M14         \\
  2& 39 &58.7 &  --34& 25& 28.1 &   1018 & 14.69 & 13.77 & 13.51 &  1.19  &02395861-3425279  &SC4/8&WEL-M11 BW49 S87  \\
  2& 39 &59.4 &  --34& 19&  0.1 &   8013 & 14.88 & 14.11 & 13.93 &  0.94  &02395941-3418596  &                    \\
  2& 39 &59.5 &  --34& 32& 43.7 &   4029 & 15.10 & 14.20 & 14.01 &  1.08  &02395951-3432432  &                    \\
  2& 40 & 0.1 &  --34& 29& 50.9 &   1030 & 15.48 & 14.62 & 14.44 &  1.04  &02400007-3429507  &                    \\
  2& 40 & 1.8 &  --34& 44&  4.4 &  34016 & 15.21 & 14.39 & 14.26 &  0.95  &02400181-3444039  &                    \\
  2& 40 & 2.7 &  --34& 48& 26.1 &  34010 & 15.35 & 14.54 & 14.23 &  1.11  &02400272-3448257  & C & DK5 BW54 S93   \\
  2& 40 & 4.1 &  --34& 20& 11.3 &   8012 & 15.18 & 14.29 & 14.06 &  1.13  &02400414-3420108  &                    \\
  2& 40 & 4.4 &  --34& 34& 40.1 &   4019 & 14.49 & 13.59 & 13.26 &  1.24  &02400435-3434397  & C & DK64           \\
  2& 40 & 5.6 &  --34& 27& 43.3 &   1002 & 14.68 & 13.72 & 13.38 &  1.29  &02400553-3427432  &SC2/8& DK54 BW56    \\
  2& 40 & 6.2 &  --34& 35& 28.1 &   4051 & 15.20 & 14.37 & 14.12 &  1.08  &02400618-3435276  & C & WEL-C6         \\
  2& 40 & 7.9 &  --34& 30& 14.4 &   1013 & 15.13 & 14.26 & 13.88 &  1.24  &02400789-3430142  & C & WEL-C11        \\
  2& 40 &11.0 &  --34& 33&  0.8 &   4059 & 15.43 & 14.57 & 14.35 &  1.08  &02401104-3433004  &                    \\
  2& 40 &11.1 &  --34& 29&  4.3 &   1034 & 15.49 & 14.66 & 14.47 &  1.02  &02401102-3429042 & & BW63              \\
  2& 40 &11.9 &  --34& 28& 54.4 &   1016 & 15.11 & 14.25 & 14.03 &  1.09  &02401186-3428544  &                    \\
  2& 40 &12.5 &  --34& 39&  6.2 &  15021 & 14.85 & 13.99 & 13.75 &  1.11  &02401246-3439057  &                    \\
  2& 40 &13.7 &  --34& 43& 23.2 &  15016 & 14.93 & 14.10 & 13.87 &  1.06  &02401368-3443228  &                    \\
  2& 40 &14.1 &  --34& 23& 31.5 &   8006 & 15.01 & 14.13 & 13.85 &  1.16  &02401403-3423316  &                    \\
  2& 40 &14.4 &  --34& 34& 12.6 &   4023 & 14.97 & 14.08 & 13.81 &  1.16  &02401438-3434122 & M & WEL-M3          \\
  2& 40 &14.6 &  --34& 30& 41.2 &   1012 & 14.87 & 13.94 & 13.68 &  1.19  &02401454-3430410 & M & WEL-M7          \\
  2& 40 &15.7 &  --34& 14& 23.2 &  23005 & 15.32 & 14.56 & 14.44 &  0.87  &02401565-3414227  &                    \\
  2& 40 &16.0 &  --34& 35& 29.7 &   4017 & 15.05 & 14.17 & 13.97 &  1.08  &02401595-3435293  &                    \\
  2& 40 &16.0 &  --34& 23& 19.4 &   8022 & 15.55 & 14.71 & 14.54 &  1.01  &02401603-3423189  &                    \\
  2& 40 &17.1 &  --34& 22& 12.2 &   8009 & 14.95 & 14.10 & 13.83 &  1.12  &02401709-3422118  &M2S& DK49 BW68 S116      \\
  2& 40 &17.9 &  --34& 38& 14.6 &  15022 & 15.51 & 14.72 & 14.56 &  0.95  &02401789-3438142  &                    \\
  2& 40 &19.0 &  --34& 18& 46.7 &   8015 & 14.71 & 13.74 & 13.48 &  1.23  &02401904-3418465  &SC3/8& WEL-M19 BW70 S119 \\
  2& 40 &19.6 &  --34& 32& 50.9 &   4061 & 15.31 & 14.51 & 14.34 &  0.97  &                  & M & WEL-M4         \\
  2& 40 &22.7 &  --34& 40& 37.7 &  15019 & 15.07 & 14.21 & 13.99 &  1.07  &02402267-3440373  & & BW73 S124 \\
  2& 40 &23.3 &  --34& 12&  5.8 &  23009 & 15.29 & 14.43 & 14.33 &  0.97  &02402332-3412058  &                    \\
  2& 40 &23.4 &  --34& 43& 23.0 &  15006 & 14.74 & 13.83 & 13.38 &  1.36  &02402342-3443225  & C & WEL-C17        \\
  2& 40 &23.4 &  --34& 34& 15.2 &   4022 & 15.07 & 14.24 & 14.02 &  1.05  &02402345-3434148  &                    \\
  2& 40 &23.5 &  --34& 15& 20.5 &  23004 & 15.25 & 14.32 & 14.20 &  1.04  &02402354-3415204  & & ET0161                   \\
  2& 40 &23.9 &  --34& 18& 10.9 &   8017 & 15.10 & 14.19 & 13.95 &  1.16  &02402389-3418108  &                    \\
  2& 40 &24.1 &  --34& 40& 57.7 &  15018 & 15.31 & 14.51 & 14.31 &  1.00  &02402411-3440572  &                    \\
  2& 40 &24.1 &  --34& 34& 18.5 &   5013 & 14.98 & 14.11 & 13.76 &  1.22  &02402415-3434182  & C & WEL-C16        \\
  2& 40 &24.5 &  --34& 13& 26.8 &  23006 & 15.34 & 14.57 & 14.42 &  0.92  &02402450-3413268  &                    \\
  2& 40 &30.8 &  --34& 22&  3.1 &   7010 & 15.49 & 14.75 & 14.52 &  0.97  &02403081-3422028  &                    \\
  2& 40 &31.7 &  --34& 27& 55.6 &   6020 & 14.90 & 14.04 & 13.85 &  1.05  &                  &                    \\
  2& 40 &32.0 &  --34& 20& 48.9 &   7012 & 14.98 & 14.09 & 13.77 &  1.21  &02403195-3420487  &                    \\
  2& 40 &34.1 &  --34& 23& 50.6 &   6028 & 15.12 & 14.26 & 13.95 &  1.17  &02403414-3423503  & C & DK51           \\
  2& 40 &36.7 &  --34& 17& 13.7 &   7015 & 15.05 & 14.22 & 13.90 &  1.15  &02403668-3417137 &  &                  \\
  2& 40 &39.0 &  --34& 26& 26.0 &   6025 & 15.53 & 14.81 & 14.66 &  0.87  &02403902-3426257  &                    \\
  2& 40 &39.0 &  --34& 26& 46.5 &   6024 & 15.35 & 14.44 & 14.24 &  1.11  &02403907-3426461  &                    \\
  2& 40 &39.6 &  --34& 36& 10.0 &   5011 & 15.10 & 14.20 & 14.04 &  1.06  &02403957-3436095  &                    \\
  2& 40 &39.8 &  --34& 31& 33.5 &   5015 & 15.18 & 14.35 & 14.11 &  1.08  &02403983-3431330  &                    \\
  2& 40 &40.0 &  --34& 29& 37.4 &   6009 & 14.97 & 14.06 & 13.82 &  1.14  &02404004-3429370  &                    \\
  2& 40 &41.1 &  --34& 23& 55.2 &   6027 & 15.43 & 14.64 & 14.32 &  1.11  &02404109-3423549  &                    \\
  2& 40 &45.8 &  --34& 37& 55.9 &  16006 & 15.06 & 14.21 & 14.10 &  0.96  &02404579-3437560  &                    \\
  2& 40 &50.0 &  --34& 25& 58.4 &   6015 & 14.76 & 13.85 & 13.53 &  1.23  &02405005-3425580  &                    \\
  2& 40 &55.3 &  --34& 27& 25.7 &   6022 & 15.33 & 14.40 & 14.18 &  1.15  &02405536-3427252  &                    \\
  2& 40 &56.8 &  --34& 27&  9.7 &   6014 & 15.11 & 14.18 & 13.95 &  1.16  &02405687-3427093  &                    \\
  2& 41 & 5.9 &  --34& 27& 32.4 &  19004 & 15.05 & 14.20 & 13.94 &  1.11  &02410593-3427318  &                    \\
  2& 41 &10.8 &  --34& 31& 51.9 &  18009 & 15.05 & 14.16 & 13.92 &  1.12  &02411082-3431518 & & BW84 S153 DI16             \\
  2& 41 &22.4 &  --34& 11& 29.9 &  21012 & 15.26 & 14.34 & 14.17 &  1.09  &02412242-3411296  &                    \\
  2& 41 &26.2 &  --34& 28&  6.0 &  19003 & 14.40 & 13.53 & 13.26 &  1.14  &02412623-3428056  &                    \\
  2& 41 &27.0 &  --34& 22&  7.7 &  20009 & 14.94 & 14.04 & 13.75 &  1.19  &02412701-3422074  &                    \\
\\
\end{longtable}
\end{center}
\normalsize
\twocolumn

Table~\ref{agb} lists all 120 of the AGB
candidates (omitting the variables listed in Tables \ref{tab_LPV} and
\ref{vars}), together with their 2MASS and other names and spectral types
when available.  The mean $JHK_S$ of these stars shows very little scatter
($\sigma<0.11$ mag), so if they are variables they have low infrared
amplitudes. Thirty of these are in fact in Bersier \& Wood's (2002) list of
candidate long-period variables, so they may have larger amplitudes at
shorter wavelengths. Nine of the AGB stars are in common with Battaglia et al.
(2006) and all are radial velocity members of Fornax. Their metallicities,
derived from CaII, fall in the range $\rm -1.01<[Fe/H]<-0.57$, with a mean
[Fe/H]=--0.86, i.e. essentially identical with the models illustrated in the
colour-magnitude diagram (Fig.~\ref{fig_cm}).

Fig.~\ref{fig_pos} shows the positions of the upper AGB candidates and AGB
variables on the sky. Note that Fornax extends well beyond the area that our
observations cover; the tidal radius is $71'\pm 4'$ (Irwin \& Hatzidimitriou
1993) and globular clusters numbers 1 and 5 fall outside our region (see
also Coleman \& de Jong's (2008) figs.~1 and 3). The overall distribution is
more or less as one would expect given the well known ellipsoidal shape of
the galaxy. It is interesting that the Miras, which are probably
representative of the intermediate age population, are not highly
concentrated towards the centre.

The Marigo et al. (2008) models, discussed above, fit these AGB stars much
better than any models published previously. One of the major improvements
is the use of proper molecular opacities as convective dredge up occurs,
another is the incorporation of dust into the radiative transfer
calculations for the AGB, and several different prescriptions for the dust
composition are allowed.  For the illustrated isochrone the dust was a
mixture of silicates (for O-rich stars) and graphites (for C-rich stars)
from Bressan et al. (1998). Using the alternative dust combinations from
Groenewegen (2006) results in an AGB that terminates at $J-K_S<3.7$ mag,
before it reaches the reddest of our AGB stars, but is closer to the
observed points in the two-colour diagram (Fig.~\ref{fig_jhhk1}). The
illustrated isochrone becomes carbon rich as it goes brighter than
$K_S=14.2$ mag. As discussed below most (although not all) of the stars
brighter than this, that have spectra, are indeed C-rich. Isochrones of 1 or
3 Gyr or with Z=0.001 and 0.005 do not pass through as many of the observed
points. Note that stars on the 10 Gyr track never become carbon rich.

The isochrone fits are discussed further in section 7 in relation to the 
long-period variable stars.

\section{Stars with spectral types}
\begin{figure} 
\includegraphics[width=8.5cm]{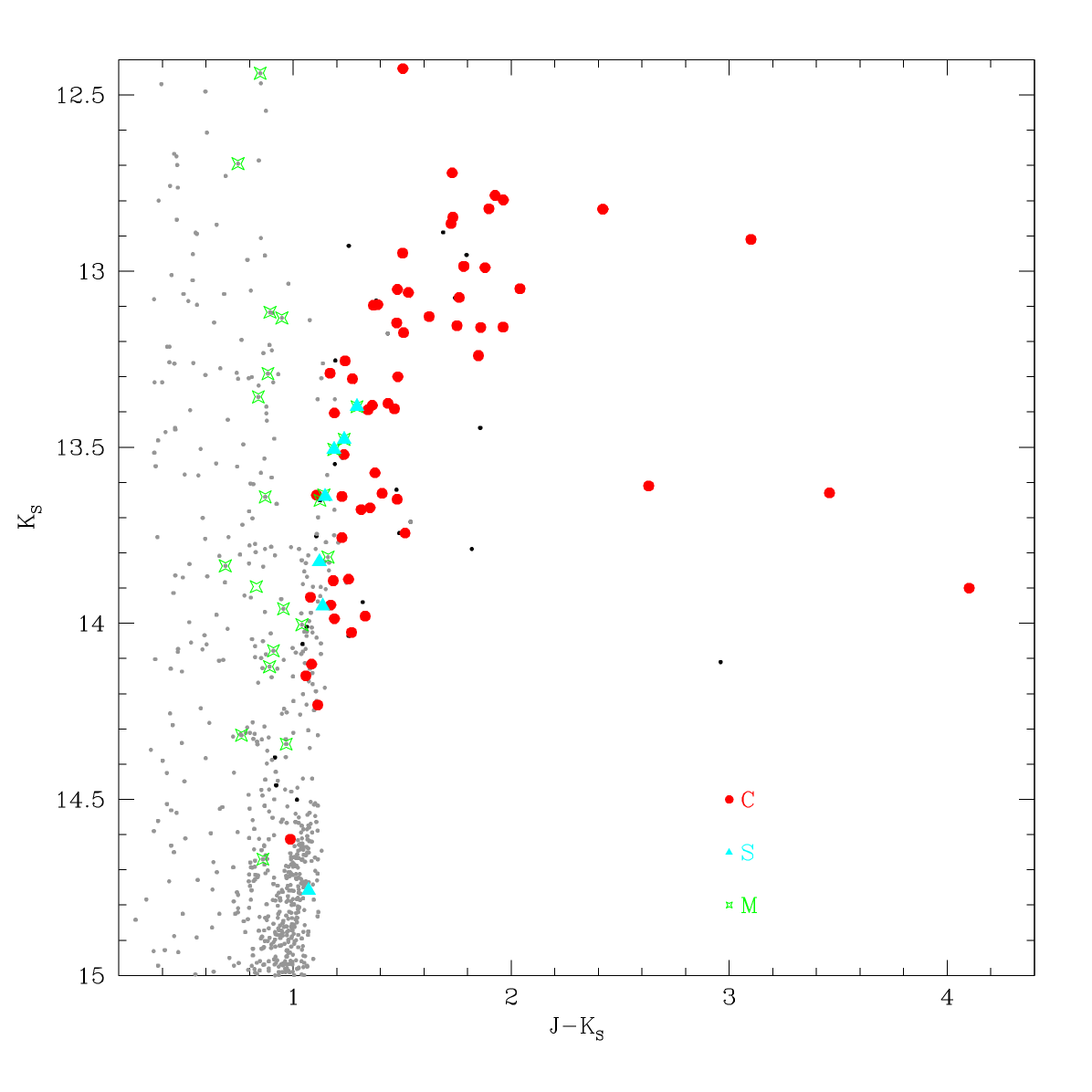}
\caption{Spectroscopically confirmed stars are illustrated as large circles
for carbon stars, triangles for S stars (including MS and SC classes) and
asterisks for M stars. }
\label{fig_sp}
\end{figure}

 An extensive spectroscopic survey of Fornax by Westerlund et al. (1987) and
by Lundgren (1990) confirmed the presence of some known carbon stars and led
to the discovery of many others. Furthermore, M-type spectra were found for
a large number of their survey stars, while S characteristics were found for
a small number. In Fig.~\ref{fig_sp} we show all the C, S and M stars from
these surveys over-plotted on our colour-magnitude diagram. 

Demers et al. (2002) examined 2MASS sources within a narrow colour-magnitude
range and identified 5 new carbon stars. Mauron et al. (2004) confirmed one
of these (their M30). Groenewegen et al. (2008) identified further C-stars,
including some of the variables discussed below, from near-infrared spectra
of 2MASS sources. Matsuura et al. (2007) observed five stars from Fornax,
selected from our survey, with the infrared spectrometer on the Spitzer
Space Telescope, demonstrating all to be C-rich. These stars are cross
identified in Tables~\ref{agb}, \ref{tab_LPV} and \ref{vars} as well as
being marked in Fig.~\ref{fig_sp}. Note that those G/K stars identified by
Groenewegen et al. (2008) that are within the area that we surveyed are all
below the TRGB.

\begin{figure}
\includegraphics[width=8.5cm]{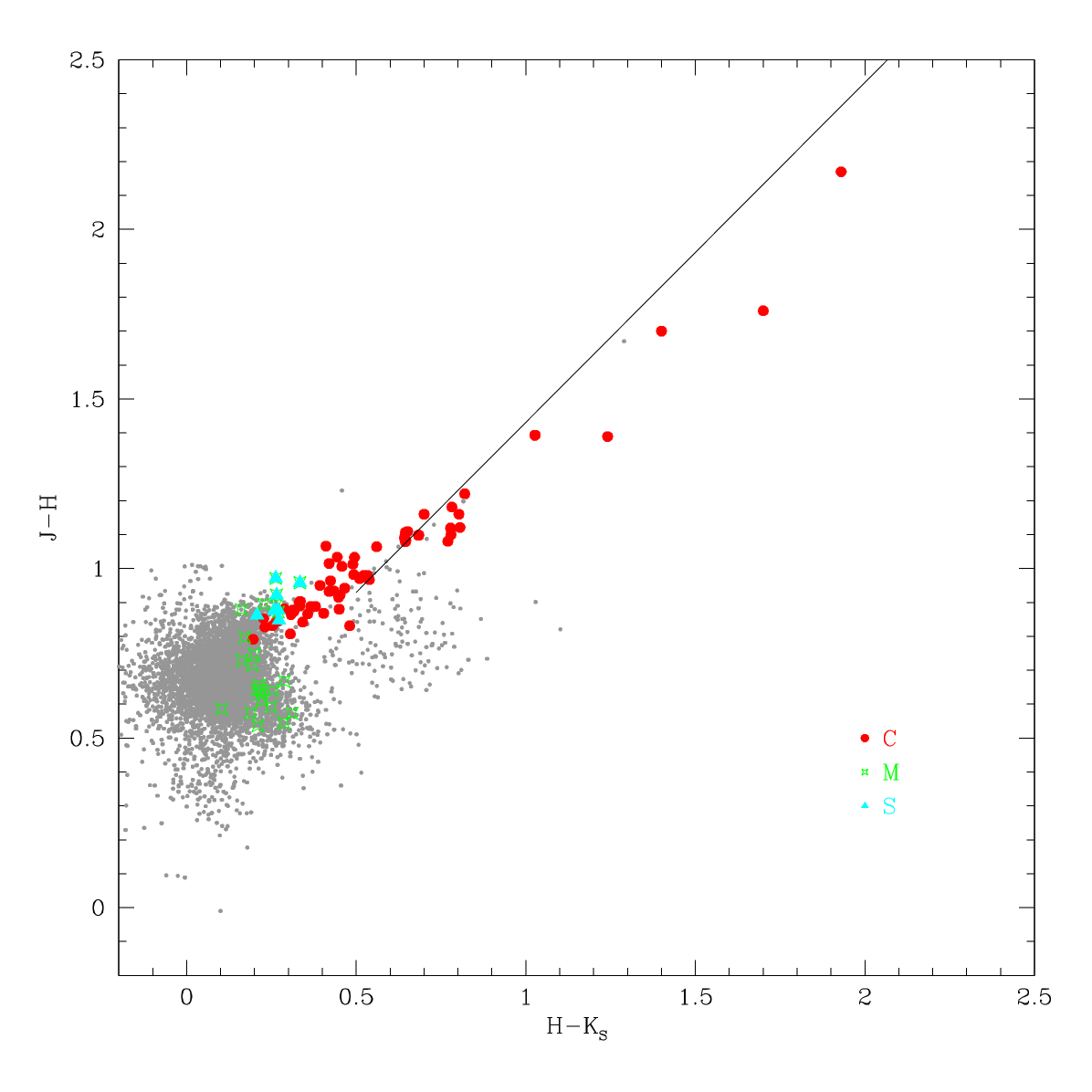}
\caption{Two colour diagram for Fornax. Symbols as in Fig.~\ref{fig_sp}. }
\label{fig_jhhk2}
\end{figure}

During the later stages of AGB evolution thermal pulsing results in the 
dredge up of carbon to the surface of the star. The surface C/O-ratio
will gradually increase, and the star will go from being O-rich (spectral
type M or S) to C-rich (spectral type C). Stars with S-type spectra have
C/O ratios less than, but very close to, one (although see Lebzelter et al.
(2008) for an exception to this). If the initial metallicity of the 
star is low it will take very little dredge-up to bring C/O to more than one
and this transition therefore tends to occur earlier in metal-weak
populations. There is also a mass dependence, in that this so called ``third
dredge-up" does not occur at all in low mass stars (Iben \& Renzini 1983). 

The transition from O-rich to C-rich thus obviously depends on the mass and
metallicity of the parent population. The luminosity at which this
transition occurs is predicted by the models and it is therefore of interest
to establish the maximum luminosity for the O-rich stars and the minimum for
C-rich stars. Within that context it is important to establish the
luminosity range over which M, C and S stars are found. Note that it is also
possible to get extrinsic C or S stars, when enriched material is deposited
onto the surface of another star to give that star the appearance of having
undergone dredge-up before it actually does so. This is generally thought to
be the explanation for low luminosity S and C stars.

There are nine spectroscopically confirmed M stars in our sample
with luminosities in the range $14.34>K_S>13.36$ mag, all with similar colours
$0.9 <J-K_S<1.2$ mag. There are also seven stars in our sample with S
characteristics (S, SC or MS spectral types). Six of these are listed in
Table~\ref{agb}, the seventh, DK54, is fainter than the TRGB and must
therefore be assumed to be extrinsic, as dredge-up is not predicted to occur
so low on the AGB.  The other six have $13.95>K_S>13.38$ and $1.1
<J-K_S<1.3$ mag. None of these M or S stars has colours suggesting significant
mass loss and their ranges in bolometric luminosity will be similar to their
ranges in $K_S$ luminosity. The mix of spectral types and luminosities among
the S stars, e.g. the faintest two are SC type and the brightest two are S
type, suggests that these stars cannot be from a single population, but must
have a range of masses and/or metallicities. However, Lebzelter et al. (2008)
who do a detailed abundance analysis of AGB stars in the intermediate age
LMC cluster, NGC\,1846, find a similar mixture of luminosity and C/O ratios
for the O-rich stars in their study. In view of the fact that NGC\,1846 is
not expected to have the large spread in ages that Fornax does (although it
is probably not a single age population) they suggest this is the result of some
of the stars being in low luminosity phases following a shell flash and
dredge-up. A similar explanation may contribute to the luminosity spread
among the S stars in Fornax.

It seems possible that some of the M stars could be in the foreground and
not members of Fornax, and some could be representatives of the younger
($<1$ Gyr) Fornax population known to be present.  It is highly unlikely
that the S stars are anything but members of the Fornax intermediate age
population.

There is a carbon star as faint as $K_S=14.6$ mag, (F23007 = WEL-C19), which is not
in Table~\ref{agb} (it is shown in Fig.~\ref{fig_sp}) as it is just fainter than
our selection criterion. It may be on the AGB or it may be extrinsic. The
rest of the C stars are brighter than $K_S \sim 14.2$ mag and there are four with
$14.2>K_S>14.0$ mag, which seems to be the lower luminosity limit for most of
the C-stars. This coincides with the predicted lower luminosity limit for
the Marigo et al. (2008) model for a population of 2\,Gyr and a metallicity of
Z=0.0025. Nevertheless, the mix of S and C stars of comparable luminosity
suggest we are sampling the AGB of a population spanning a range of mass
and/or metallicity.

\section{Variable stars}
\begin{figure}
\includegraphics[width=8.5cm]{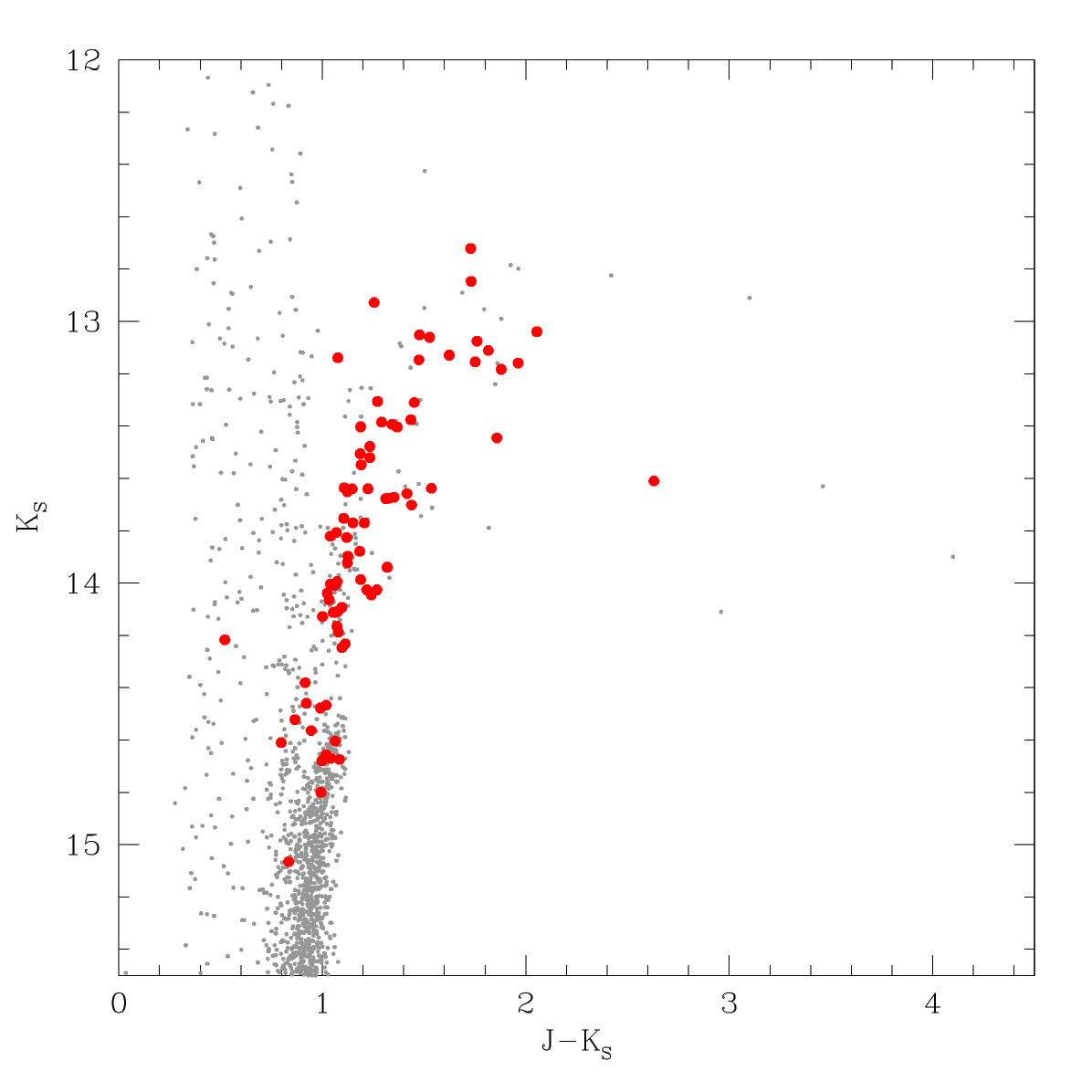}
\caption{Colour-magnitude diagram for Fornax. 
Large filled circles show stars in common with Bersier \& Wood (2002).}
\label{fig_cmBW}
\end{figure}

An extensive search for variable stars in Fornax by Bersier \& Wood (2002)
(hereafter BW) yielded many RR Lyrae stars, anomalous Cepheids and
Population II Cepheids, together with 85 candidate long-period variables. We
have cross-identified the candidate long-period variables (8 lie outside our
survey area) with our catalogue and display the result in
Fig.~\ref{fig_cmBW}. It seems that most stars on the upper AGB, as well as
some at the tip of the giant branch are variable. These latter stars are
probably very similar to those found by Ita et al. (2002) in the LMC. The
variables identified in the present study are listed in Tables~\ref{tab_LPV}
and ~\ref{vars}. For those in the latter table, it was not possible to
derive periods, but the observed peak to valley range is given ($\delta J,\
\delta H,\ \delta K_S$). Potential variables were required to have a range
of at least 0.1 mag and to show similar behaviour at each epoch in each
waveband.

\begin{figure}
\includegraphics[width=8.5cm]{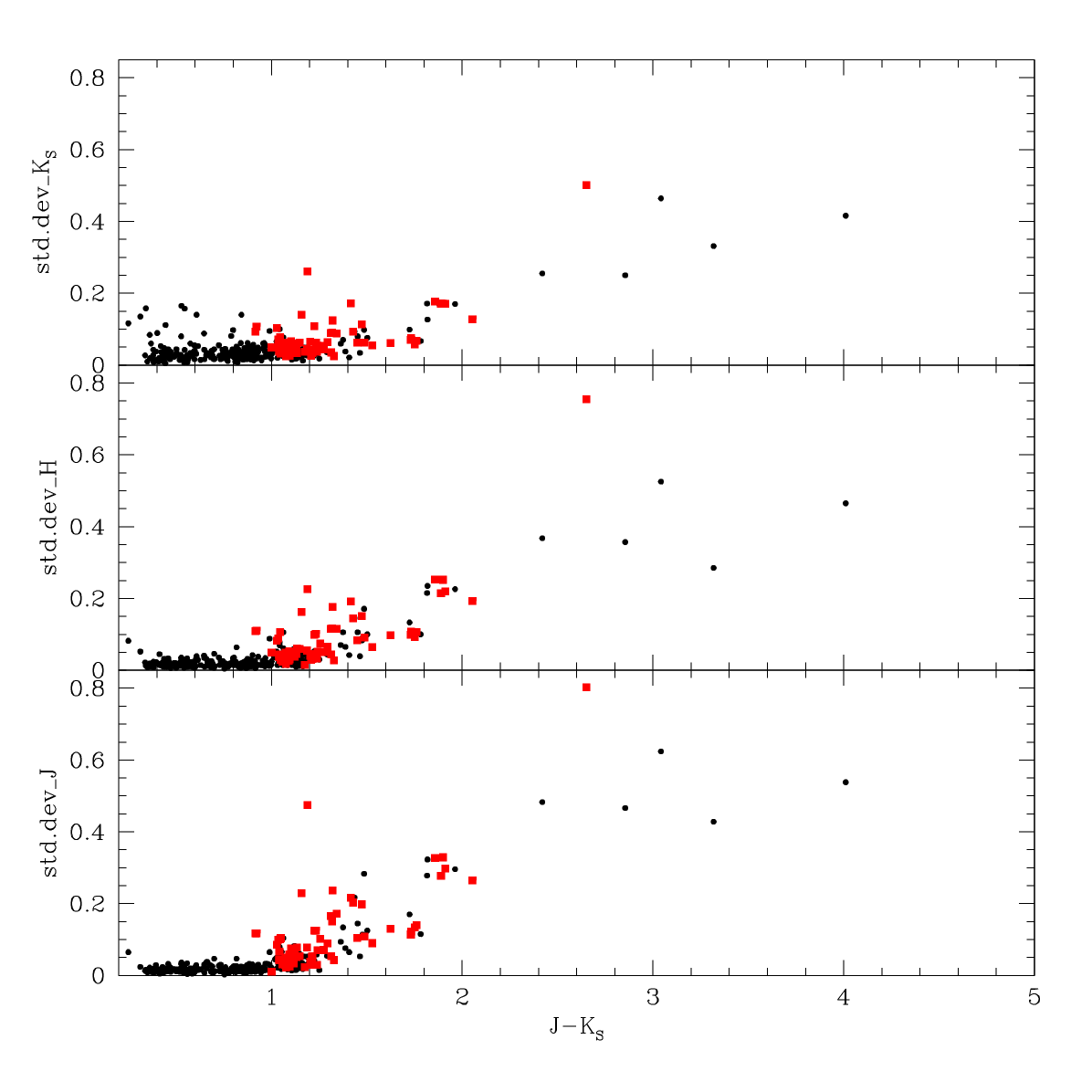}
\caption{Standard deviation in $JHK_S$ as a function of $J-K_S$ colour.
Squares denote stars in common with the variable star list of Bersier \& Wood (2002). Note the increase in
standard deviation towards redder colours, interpreted here as due to
increasing amplitude of variability.}
\label{fig_sdevjk}
\end{figure}

It is worth checking that the stars in common with BW are indeed variable in
the infrared. As a first step, we show in Fig.~\ref{fig_sdevjk} the
standard deviation for stars brighter than $K_S=14.5$ mag as a function of
$(J-K_S)$. Until about $(J-K_S)=1.2$ mag, the standard deviation is approximately
constant, and thereafter increases steadily in all bands indicating
variability with increasing amplitude.

\section{Long-period Variables}

\begin{table*}
\begin{center}
\caption{Periodic Red Variables.}
\label{tab_LPV}
\begin{tabular}{rcccccccccccl}
\hline
\multicolumn{1}{c}{F} & \multicolumn{1}{c}{P} & \multicolumn{1}{c}{$J$} &  
\multicolumn{1}{c}{$\Delta J$} & \multicolumn{1}{c}{$H$} &
\multicolumn{1}{c}{$\Delta H$} &
\multicolumn{1}{c}{$K_S$}& \multicolumn{1}{c}{$\Delta K_S$} & $m_{bol}$& 
$M_{bol}$ & Sp & 2MASS&other names\\
\multicolumn{3}{c}{(day)}&\multicolumn{7}{c}{(mag)} \\
\hline
{\bf Miras}\\
 5010 & 215 &15.19 &0.62 & 14.11 & 0.54 &  13.62&  0.56 & 16.98 & --3.72&-&02405224--3437230& BW82 DK23   \\
11020 & 258 &17.07 &1.18 & 15.40 & 0.96 &  14.11&  0.74 & 17.27 & --3.43&-&02391340--3425421\\  
 1003 & 267
 &15.02 &0.87 & 13.86 & 0.65 &  13.16&  0.42 & 16.64 & --4.06&C&02400252--3427426& DK55 S92 BW52\\
36008 & 280 &14.87 &0.69 & 13.77 & 0.56 &  12.99&  0.52 & 16.48 & --4.22&C&02410355--3448053&BW83 DDB26 GLM27 M30\\
13023 & 350 &17.09 &1.28 & 15.33 & 1.24 &  13.63&  1.02 & 16.30 & --4.40&C&02385056--3440319&(13-23) GLM34 \\
 3099 & 400 &18.00 &1.64 & 15.83 & 1.43 &  13.90&  1.16 & 16.06 & --4.63&C&02394160--3435567&(3-129)\\
12010 & 470 &16.01 &1.44 & 14.31 & 1.14 &  12.91&  0.96 & 15.95 & --4.75&C&02391232--3432450&(12-4) GLM25 \\
{\bf SRs}\\
 4057 & 171 &15.31 &0.58 & 14.43 & 0.44 &  13.98&  0.30 & 17.26 & --3.44&C&02401994--3433097& WEL-C9 S122 DI19 BW72\\
22006 & 230 &15.09 &0.85 & 14.01 & 0.60 &  13.24&  0.31 & 16.72 & --3.98&C&02405333--3412130& DDB24 GLM11 \\ 
 1001 & 235 &14.78 &0.47 & 13.81 & 0.38 &  13.30&  0.27 & 16.66 & --4.04&C&02402504--3428583& S131 DK56 BW74 WEL201 \\
 3005 & 255 &14.86 &0.34 & 13.65 & 0.29 & 12.94& 0.22 & 16.42 & --4.27&C&02393738--3436268& WEL-C1\\ 
 3006 & 284 &14.68 &0.48 & 13.50 & 0.37 & 12.88& 0.28 & 16.33 & --4.37&C&02394845--3435078&\\ 
 4032 & 303 &14.92 &0.36 & 13.71 & 0.27 & 13.07&0.17  & 16.53 & --4.17&C&02400274--3431489&DK62 BW53 \\ 
4025 & 320 &15.09 &0.64 & 13.87 & 0.43 &  13.05&  0.26 & 16.54 &--4.16&C&02401016--3433218&(4-25) WEL-C10 S105 DI20 BW62 \\
\multicolumn{2}{l}{\bf {SRs+trend}}\\
32007 & 255 &15.30 & & 14.17 & &  13.45 & & &&-& 02385309--3449199& DK8 BW15  \\
1105  & 340 &16.24 & & 14.85 & &  13.61 & & &&C& 02401778--3427357& GLM21  BW69 \\
1006  & 375 &15.24 & & 13.85 & &  12.82 & & &&C&02401207--3426255& WEL-C12 GLM20 \\
\hline
\end{tabular}
\end{center}
\flushleft{Notes: other names follow the same convention as Table~\ref{agb}
while numbers in brackets indicate a name from Matsuura et al. (2007)
and numbers preceded by M are from Mauron et al. (2004).}
\end{table*}

\begin{figure}
\includegraphics[width=8.5cm]{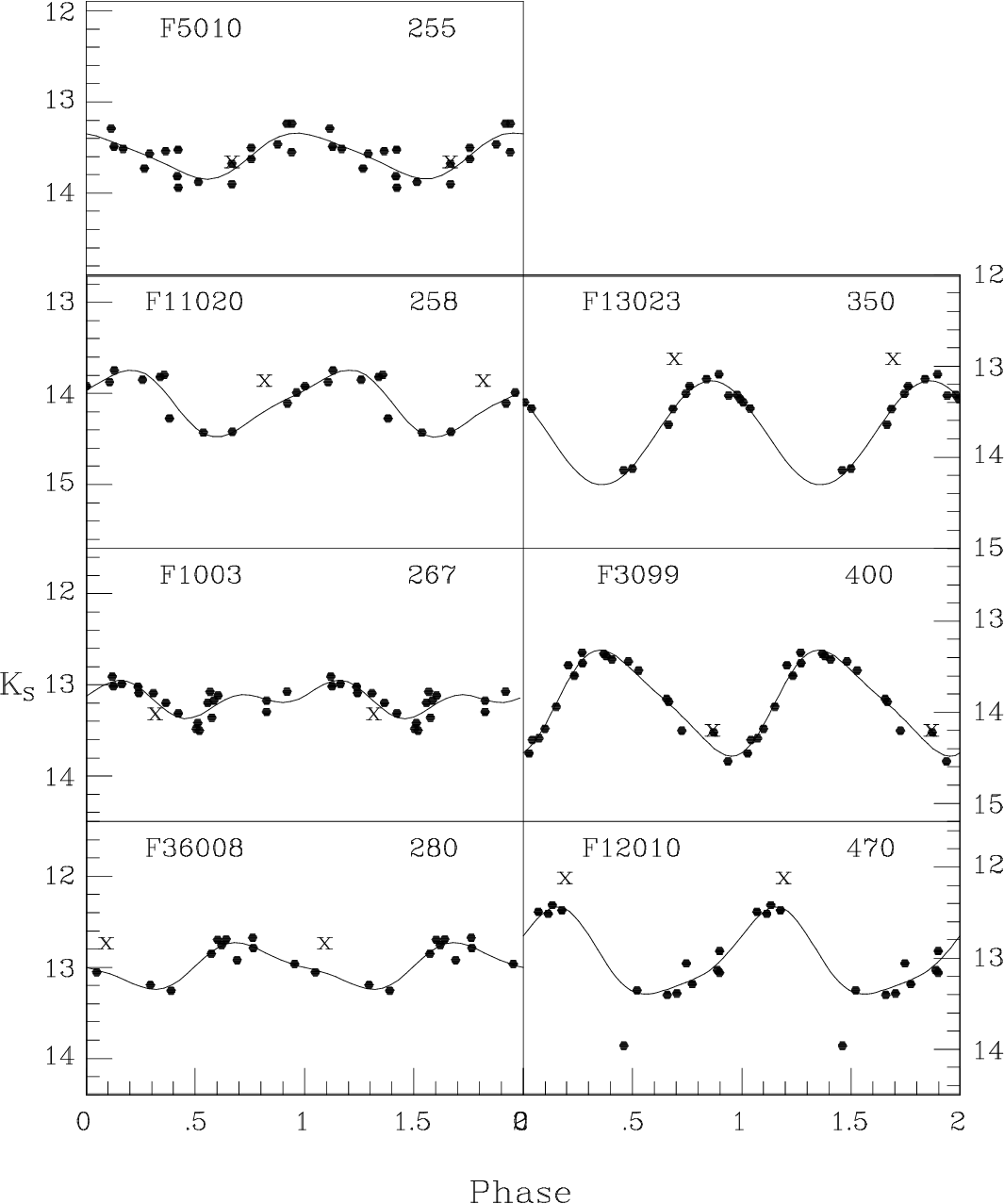}
\caption{$K_S$ light curves for the Mira variables. Each point is plotted
twice to emphasize the variability and the 2MASS photometry is shown as a
cross. The curves show the best fitting second order sine waves.}              
\label{fig_lc}
\end{figure}

Table~\ref{tab_LPV} lists the Fourier mean $JHK_S$ magnitudes for the
periodic variables together with the peak-to-peak amplitudes ($\Delta J,\
\Delta H,\ \Delta K_S$) of the best fitting second-order sine curves (i.e.
the period determined from a Fourier analysis plus its first harmonic -- this
because most of the curves show asymmetries between the rising and falling
branches). 
The $K_S$ light curves of the Miras are illustrated in
Fig.~\ref{fig_lc}, from which it is clear that some of the periods are more
regular than others. Fig.~\ref{fig_lc} also shows the 2MASS photometry,
which was obtained from observations on JD2451147, about 1000 days before
the first of our IRSF measurements.  We also attempted to redetermine the
periods using the combined data set, but the results were unsatisfactory.
The fact that some of the 2MASS points do not lie on the light curves is not
in itself surprising. The periods are only good to between 5 and 10 percent
so the phase information will not be perfect after a gap of over 1000 days
(3 or 4 cycles). In addition, while the periods of most Miras are very
stable the light curves do not repeat exactly from one cycle to the next,
and erratic changes and long term trends are a feature of C-rich Miras with
dust shells (Whitelock et al. 2006).  All of the 2MASS $K_S$ measures are good
quality with no indication of confusion.

For F12010 we calculated the period and other parameters without the last
observation, which is fainter and redder than any of the earlier values. The
2MASS observation, which predates the IRSF ones by over 1300 days, is
brighter and bluer than the IRSF observations. These could be manifestations
of a long term trend, but it is impossible to be certain without many more
observations.
 
Following Whitelock et al. (2006), who discussed infrared photometry for
Galactic carbon stars, we consider Miras to be stars with peak-to-peak
amplitudes at $K$ in excess of 0.4 mag, and for which periods can be determined.
It is on this basis that the stars in Table~\ref{tab_LPV} were divided into 
Miras and semiregulars (SRs). The total, peak-to-peak variation of the stars described 
as ``SRs+trend" is large, certainly $\Delta K_S> 0.4$ mag, but the
pulsation amplitude is relatively low. They are discussed further in section
7.1.

It seems likely that our survey for Mira variables is complete for the area
which we covered. The limiting magnitude on the $H$ reference frames is
greater than 19 mag. An AGB star in Fornax with an extremely thick shell
would probably have $H-K = 2.5$ or redder.  At the limiting $H$ magnitude
this would correspond to $K = 16.5$ mag, well above the
$K_S$ limit of about 18 mag. No stars were found on any of the $K_S$ frames
that did not have a counterpart on an $H$ frame, so it is unlikely that any
extreme Mira has been missed. Furthermore, Hinako Fukushi cross identified
objects we discuss here with those detected by AKARI (with the IRC camera in
4 bands between 3 and 11 $\mu$m) in 500 square arcmin of the area we survey.
She found nothing beyond the sources detected by the IRSF.

It is, however, likely that there are some Miras outside of the area that we
survey.  A search for 2MASS sources with $J-K_S>2$ over an area with a
radius of one degree centred on the Fornax dSph reveals two sources
outside our survey area: 02400946--3406256 and 02380618--3431194. Both of
these have been observed by Groenewegen et al. (2008), their sources
Fornax15 and Fornax31, respectively, and both are carbon stars. These two
are very likely to be Miras and there could easily be one or two other
Miras in the same region with $J-K_S<2$.

The 2 Gyr isochrone, which passes through most of the variable stars in
Fig.~\ref{fig_cm}, is probably reasonably representative of the bulk of the
Fornax AGB population.  However, the fact that this isochrone passes close
to most of the red variables should not be taken to indicate that all the
stars in this plot were formed at the same time. Indeed, given the range of
ages and metallicities that are known in Fornax, it would be extraordinary
if all the large amplitude variables were from the same population.  In the
case of oxygen-rich Miras it is known that the pulsation period is a
function of mass, as well as, at the shorter periods, of metallicity (Feast
\& Whitelock 2000). One would reasonably expect the carbon Miras to follow
parallel trends and the available evidence is consistent with this. Thus the
three Magellanic Cloud clusters discussed by Nishida et al. (2000) have ages
of about 1.6 Gyr (Mucciarelli et al. 2007a,b; Glatt et al. 2008) and their
three carbon Miras have a mean period of 490 days, whilst van Loon et al.
(2003) suggest that the cluster KMHK 1603 has an age of 0.9-1.0 Gyr and
contains a carbon Mira of period 680 days. The kinematics of Galactic carbon
Miras (Feast et al. 2006) is also consistent with a decreasing age with
increasing period. 

So far as the isochrones are concerned it is worth noting that the sequence
of stars in Fig.~\ref{fig_cm} is not a sequence of increasing period with
colour such as might be expected if the Miras were evolving with period
along a single evolutionary track (which would be essentially identical to
an isochrone for AGB stars which evolve very rapidly). Furthermore a
comparison of the bolometric luminosities of the Fornax Miras with the
models in diagrams such as the $\log L - \log T_{eff}$ plot of Marigo et al.
(their fig. 1) shows that whilst the fainter, shorter period, stars must be
amongst the oldest (lowest initial mass) carbon stars allowed by these
models, the age of the brighter, longer period carbon Miras cannot be
estimated in this way since it depends on $\log T_{eff}$ which is known very
imprecisely if at all.

The pulsation periods that the Marigo et al. (2008) models associate with
the AGB variables bear no relation to the values actually found here. 
However, stars at this stage evolve rapidly and the isochrones provided by
the on-line model go from
$J-K_S=1.6$ (P=162) to $J-K_S=8.1$ (P=257) in a single step, so it isn't
practical to match them with the detailed observations.  In theory one might
expect the pulsation period of a Mira to change as the star evolves and
loses mass. In practice the observational evidence, discussed in the
previous paragraph, suggests that the period of a Mira depends on its
initial mass and does not change significantly with time.  This is primarily
because a star only becomes a Mira briefly (lifetimes are around $10^5$ yr)
at the end of its AGB evolution. It remains possible that the Mira will
evolve very rapidly through longer periods before completely leaving the
AGB, but that has not been established.

The pulsation periods for the Miras in Fornax, and a comparison with
Magellanic Cloud clusters mentioned above, suggest ages of a few Gyr for
most of them. The short period ones are probably somewhat older and it is
possible that the 215 day Mira, F5010, could be significantly older, around
10 Gyr (it is the bluest of the Miras in Fig.~\ref{fig_cm}), but then it
would not be expected to be a carbon star.  For stars as blue as this one,
$J-K_S=1.57$, the colours do not allow us to distinguish between O- and
C-rich stars, so spectral classification for F5010 would be particularly
interesting.

In the following discussion the magnitudes and colours of the Fornax
variables are transformed to the SAAO system to allow a comparison with 
data for the Galaxy and the LMC, and also to apply the colour dependent 
bolometric corrections (Whitelock et al. 2006). We use Carpenter's (2001)
conversion between the 2MASS and SAAO systems, and specifically the
expressions given on the
web\footnote{www.astro.caltech.edu/$\sim$jmc/2mass/v3/transformations/}.

\begin{table*}
\begin{center}
\caption{Variables without periods.}
\label{vars}
\begin{tabular}{rcccccccccl}
\hline
F& \multicolumn{1}{c}{$J$} &  ${\delta J}$ &
\multicolumn{1}{c}{$H$} & ${\delta H}$ &
\multicolumn{1}{c}{$K_S$}& ${\delta K_S}$& $J-K_S$& Sp & 2MASS&other names\\
\hline
 28016 & 15.07 &0.28& 14.26&0.30 & 14.01 &0.17&  1.06 &  &02381712--3425482  & BW4 S3 DI13               \\
 30012& 15.26&0.51 & 14.45&0.36&  13.94& 0.12 & 1.32&   & 02382266--3438040  & BW7 DK15               \\            
 29013 & 15.10 &0.28& 14.31&0.27 & 14.06 &0.32&  1.04 &  &02384015--3432254  & BW11 S5              \\
 32004& 14.91&0.56 & 13.80&0.37&  13.16& 0.19 & 1.75& C & 02385700--3446340  & BW17 DI28 GLM32       \\
 32010& 14.86& 0.46& 13.97&0.33&  13.64& 0.22 & 1.22&  C& 02385704--3447489& DK7 DI29   BW18      \\
 12011 & 14.62 &0.32& 13.64&0.27 & 13.15 &0.20&  1.48 & C&02385829--3432117  &DK36 BW19 S11       \\
 25006& 15.23&0.94 & 14.23&0.57&  13.74& 0.25 & 1.49&   & 02391532--3415083&                   \\
  2021 & 14.77 &0.20& 13.88&0.17 & 13.65 &0.12&  1.12 & M&02392292--3428098  & DK38 BW28 S29\\
  2023 & 14.86 &0.20& 14.00&0.18 & 13.75 &0.16&  1.11 &  &02392984--3427232  & BW36 S39 DI14\\
  3015& 14.61&0.46 & 13.72&0.38&  13.18& 0.25 & 1.43&  C& 02393179--3436399& DK19              \\
  3001 & 14.46 &0.36& 13.54&0.27 & 13.08 &0.18&  1.38 & C&02393440--3437140  & DK20             \\
  2012& 15.25&0.45 & 14.26&0.34&  13.71& 0.28 & 1.54&  C& 02393670--3430247& WEL-C18           \\
  2026& 14.95&0.44 & 14.03&0.36&  13.57&0.26  & 1.37&  C& 02393943--3424578& DK42              \\
  9006& 14.75& 0.45& 13.85&0.30&  13.52&0.19  & 1.23&  C& 02393966--3419522& DK47 BW38 S55    \\
  9005& 13.93& 0.44& 12.92&0.38&  12.42&0.29  & 1.50&  C& 02394061--3420148& DK46              \\
  3028 & 14.58 &0.26& 13.71&0.18 & 13.31 &0.14&  1.27 & C&02394916--3431239  & DK37 BW44       \\
  3009& 14.45&0.35 & 13.37&0.31&  12.72&0.16  & 1.73&  C& 02395186--3433201& WEL-C5 BW46 S74      \\
 14017& 15.61&0.90 & 14.52&0.70&  13.79&0.25  & 1.82&   & 02395389--3444028&       \\
  4020 & 14.45 &0.13& 13.57&0.08 & 13.25 &0.08&  1.19 & C&02395398--3434244  & WEL-C4         \\
 14006& 14.71&1.00 & 13.59&0.76&  12.78& 0.43 & 1.93&  C& 02395421--3438368& GLM24   \\
  2011 & 14.45 &0.30& 13.47&0.24 & 12.95 &0.24&  1.50 & C&02395520--3425264  & WEL-C13        \\
  4013 & 14.48 &0.28& 13.52&0.22 & 13.10 &0.12&  1.39 & C&02395837--3436215  & DK22           \\
  8004& 14.53&0.36 & 13.50&0.28&  13.05&0.22  & 1.48&  C& 02400093--3422432& DK44 BW50 S89        \\
  8011& 14.81&0.66 & 13.80&0.45&  13.38& 0.24 & 1.43&  C& 02400135--3420189& DK48 BW51 DI7 S90        \\
  4009& 14.58&0.37 & 13.49&0.34&  12.85& 0.30 & 1.73&  C& 02400538--3432182& DK61 BW55 S98        \\
  8030& 15.38&0.30 & 14.63&0.24&  14.46&0.32  & 0.92&   & 02400650--3420131&      BW58 S101        \\
  1010& 14.76&0.61 & 13.60&0.52&  12.80& 0.32 & 1.96&  C& 02400666--3423222& GLM13 DK52         \\
  1017 & 14.74 &0.19& 13.83&0.14 & 13.55 &0.11&  1.19 &  &02400819--3426056  & BW59 S103          \\
  4006 & 14.59 &0.32& 13.56&0.16 & 13.06 &0.20&  1.53 & C&02400909--3434398  & WEL-C7 BW61 S104    \\
  4028& 14.74& 0.35& 13.79&0.28&  13.39&0.21  & 1.34&  C& 02401178--3432455& DK60 BW64 S108    \\
  1014 & 15.29 &0.62& 14.41&0.36 & 14.04 &0.23&  1.25 & C&02401224--3430093  & DK59 BW65 S109       \\
  4024& 14.75& 0.48& 13.69&0.37&  13.13&0.20  & 1.62&  C& 02401562--3434030& DK65 BW67 S115      \\
  4018 & 14.86 &0.18& 13.85&0.11 & 13.39 &0.10&  1.46 & C&02402037--3434548  & WEL-C8         \\
  6013& 15.12&1.00 & 13.94&0.75&  13.16& 0.51 & 1.96&  C& 02403123--3428441& (6-13) GLM17 BW75   \\
  5012& 15.30&0.42 & 14.56&0.39&  14.38&0.27  & 0.92&   & 02404595--3434542&      BW80 DI22        \\
  5002 & 14.18 &0.38& 13.26&0.26 & 12.93 &0.11&  1.25 &  &02404839--3435364  & BW81 S144          \\
  5016& 15.52&0.58 & 14.80&0.34&  14.50& 0.26 & 1.02&   & 02404870--3430533&                   \\
 18008& 14.84&0.44 & 13.88&0.34&  13.39& 0.20 & 1.45&   & 02410830--3435525&                   \\
\hline
\end{tabular}
\end{center}
\flushleft
Notes: the other names in the last column follow the same convention as
Tables~\ref{agb} and \ref{tab_LPV}.
\end{table*}

\subsection{Long Term Trends}
 Whitelock et al. (2003) identified C-Miras in the LMC which underwent
variations on very long time-scales, while Whitelock et al. (2006) identified
both Miras and irregular or semi-regular Galactic variables which behaved similarly 
(see figs.~18 and 19 of Whitelock et al. 2006).  They suggested
that the phenomenon is similar to that seen among the hydrogen-deficient RCB
stars and is the result of the erratic ejection of puffs of dust in random
directions. When a puff is ejected into our line of site the star fades, but
the pulsation continues. Two dimensional models for dust-driven winds in AGB
stars (Woitke 2006) suggest this is a very plausible explanation.

\begin{figure}
\includegraphics[width=8.5cm]{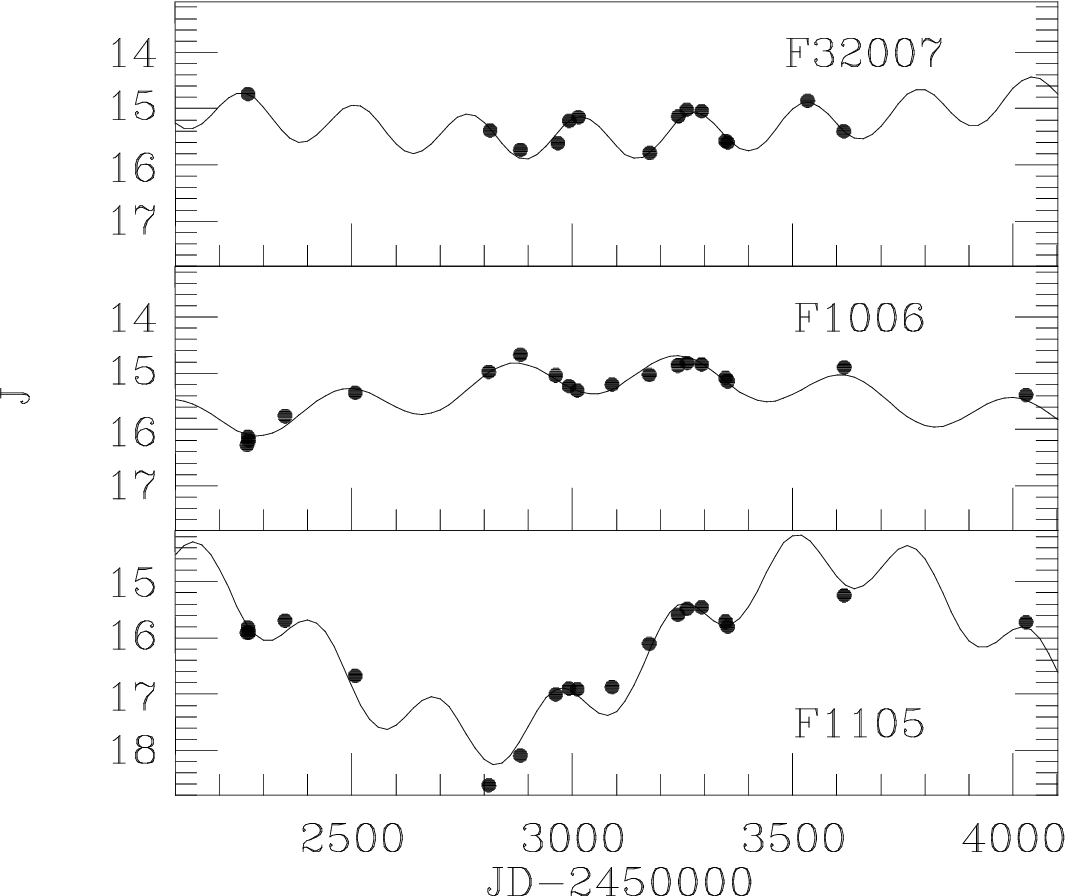}
\caption{$J$ light curves for SR variables with long term trends. 
For F32007 the line is a
combination of sine curves with periods of 255 and 3000 days,              
for F1006 the sine curves have periods of 380 and 2000 days
and for F1105 they are 270 and 1600 days, the two faint points on this curve
have higher than usual errors, but still less than 0.2 mag.}
\label{fig_jd1}
\end{figure}

Fig. \ref{fig_jd1} shows the $J$ light
curves of the three Fornax C-stars which show a combination of SR variations
and long term trend. The figure caption indicate the combination of periods
that are used in the illustrated fit. The shorter of the two periods for
each star has an amplitude $K_S<0.4$ mag, so we would classify them as SR
rather than Mira variables. The longer periods will almost certainly not
persist and are simply used here as a tool that allows us to model the trend
over a relatively short time and see the underlying periodicity.

\begin{figure}
\includegraphics[width=8.5cm]{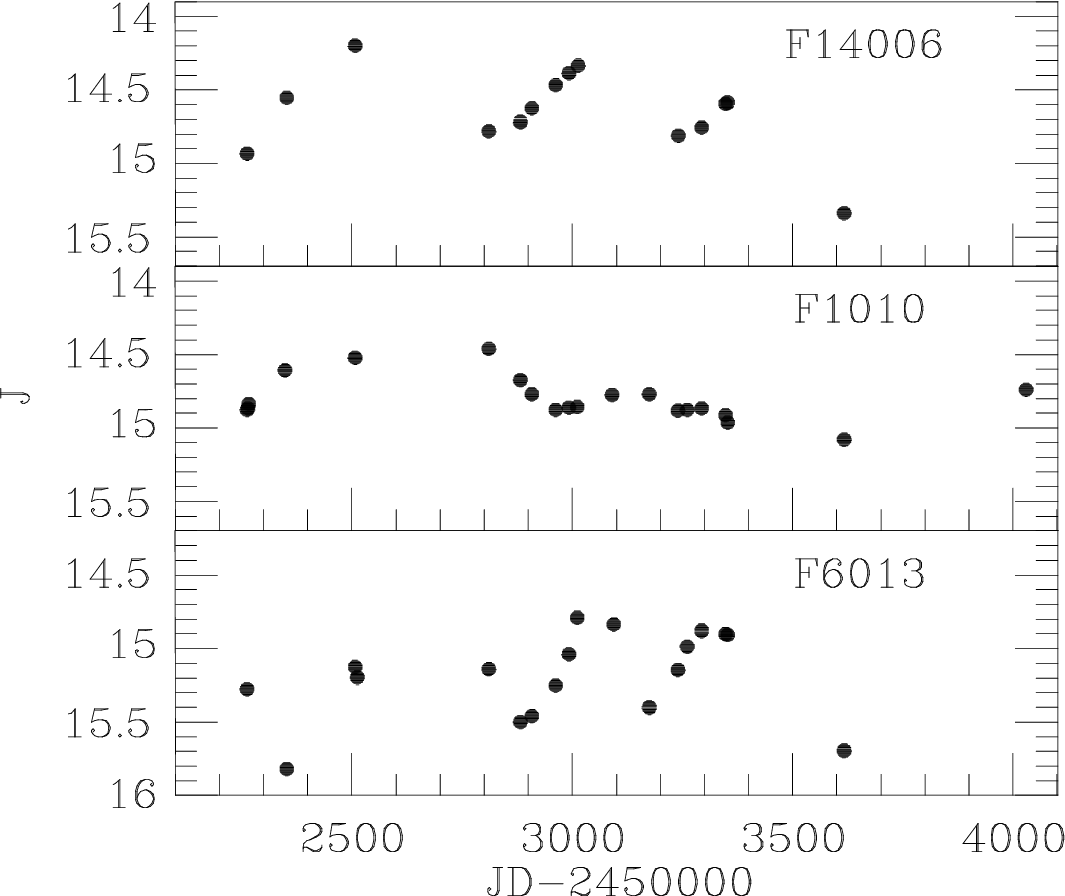}
\caption{$J$ light curves for three of the large amplitude non-periodic
variables from Table~\ref{vars}.}
\label{fig_vars}
\end{figure}

Figs. \ref{fig_vars} show the $J$ light curves of three of the largest
amplitude variables from Table~\ref{vars}; all are clearly variable. It is
possible that F6013, F1010 and F14006, which are all carbon-stars and all
rather red ($J-K_S>1.9$), are Miras or SRs undergoing extremely erratic
mass-loss, such as occurs in the very well studied Galactic carbon Mira, R
For, from time to time (Whitelock et al. 1997) and if monitored for long
enough would show periodicity and light curves more like those of Fig.~
Fig.~\ref{fig_jd1}. 

If we are correct in assuming that the Mira, F11020, is
faint in the period luminosity (PL) diagram (see section 7.3) because it is
undergoing an obscuration event then it is exhibiting the same phenomenon
and we can expect it to brighten at some future time as the obscuring dust
expands and becomes optically thin.

\subsection{Colours}
\begin{figure}
\includegraphics[width=8.5cm]{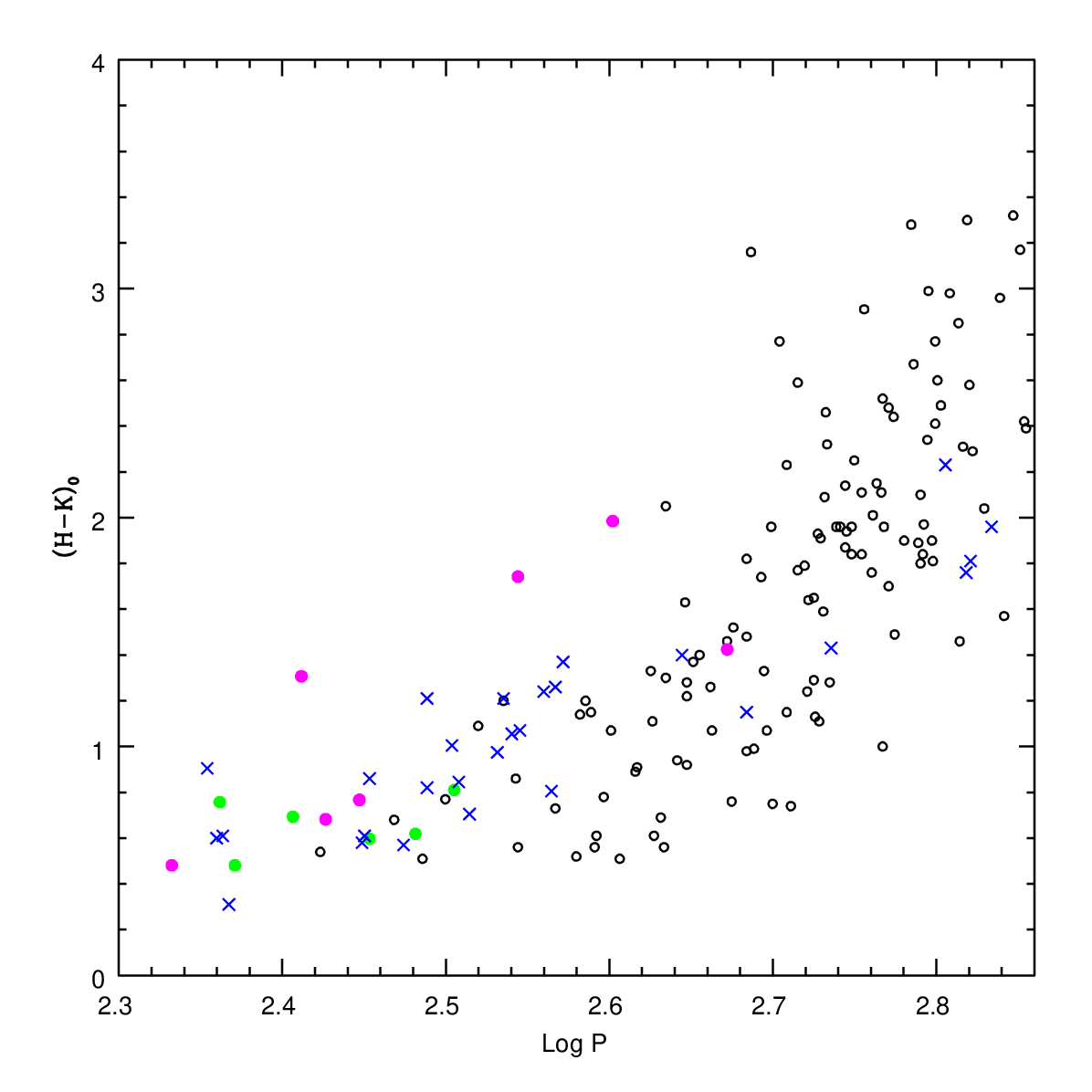}
\caption{Period-colour plot showing the Fornax Miras (solid circles) and
SRs (triangles) relative to Galactic Miras (open circles) and LMC 
Miras (crosses). All colours are on the SAAO system and have been corrected
for interstellar reddening which is significant for some of the Galactic
sources.}
\label{fig_hkp}
\end{figure}

 Fig.~\ref{fig_hkp} shows the $(H-K)$ colour as a function of period for the
periodic variables in Fornax. A comparison is provided with the colours of
C-rich Miras in the Galaxy (Whitelock et al. 2006) and in the LMC (Feast et
al. 1989; Whitelock et al. 2003).  While there is a large range of colours
at a given period within each galaxy some trends are apparent and the
colour-period distribution differs amongst these three galaxies. The LMC
stars are, on average, redder than the Galactic ones at short periods and
bluer at very long periods (but see below). The Fornax stars are redder than
their LMC counterparts at short period; there are no Miras in Fornax with
periods above 500 days.

Ita et al. (2004) discuss the characteristics of over 8000 LMC variables
with $JHK_S$ photometry from the SIRIUS camera and periods from OGLE-II. 
Their on-line data can be used to show that the LMC Miras (selected
with $\Delta I > 0.9$ mag) overlap the colours of the Galactic sample, so that
much of the separation seen in Fig.~\ref{fig_hkp} must be due to the
selection effects in the samples studied from the different locations. The
three Fornax stars which stand out in Fig.~\ref{fig_hkp} as redder than
other stars with the same period range are at the upper envelope of
colours of the LMC Miras from the Ita et al. sample, i.e. even with a much
larger sample the proportion of very red stars in Fornax is greater than it
is in the LMC.

The extreme colours are obviously due to thick dust shells as a consequence
of high mass-loss rates. Matsuura et al. (2007) note that two (F13023 and
F3099) of these three red stars also have particularly strong acetylene
bands at 7.5 $\mu$m (the third one was not measured) for their colour,
indicating strong line blanketing. The longer period Mira (F12010), which
has bluer colours, has weaker acetylene absorption, despite having a
mass-loss rate that is comparable to the two redder stars. Mass-loss rates
among carbon stars are largely independent of metallicity, but strongly
dependent on the C/O ratio, as that is what determines the amount of carbon
that is free to make dust (e.g. Lagadec \& Zijlstra 2008, but see van Loon
et al. (2008) for a different view). It may be that some of the differences
we observe between the Fornax AGB variables are due to differences in C/O,
e.g. because of differences in dredge-up, but considerably more work will be
required to clarify this.

\subsection{Bolometric Magnitudes and the PL relation}
 
\begin{figure}
\includegraphics[width=8.5cm]{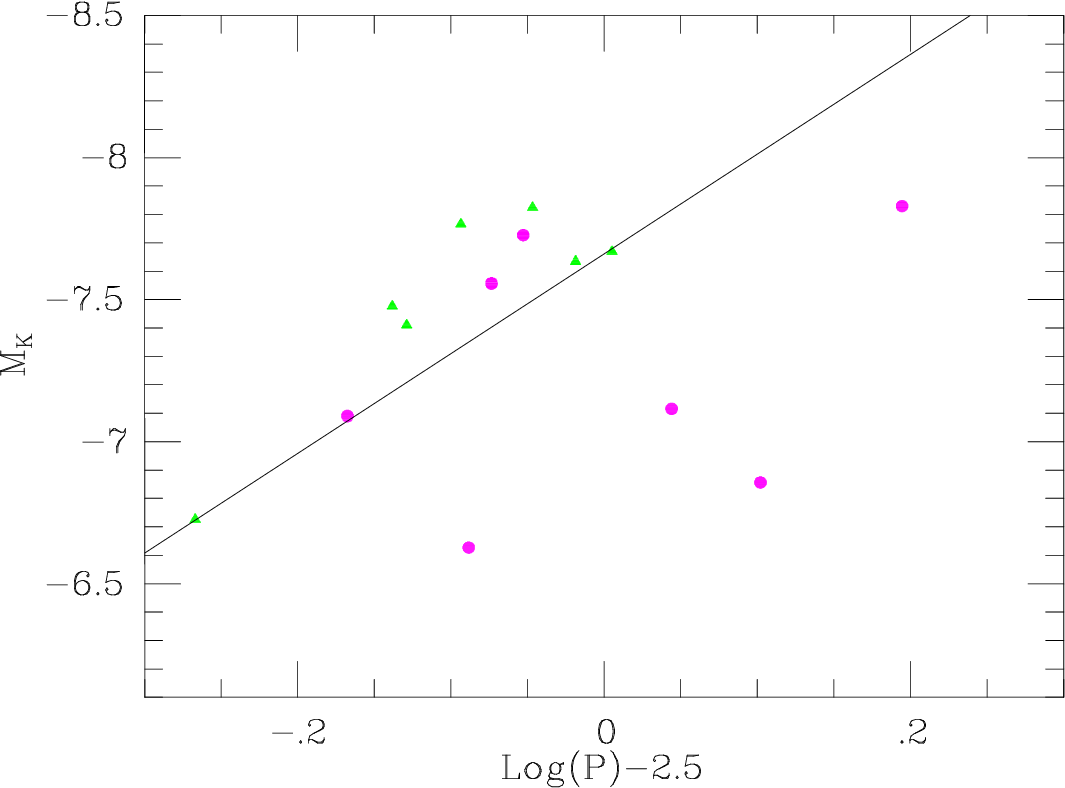}
\caption{The $M_K$ PL relation for the periodic variables in Fornax
on the assumption that the distance modulus is 20.69 mag;
Miras are shown as circles and SRs as triangles. The line is the $K$ PL
relation derived from LMC Miras.}
\label{fig_kpl}
\end{figure}

Fig.~\ref{fig_kpl} shows the $K$ PL relation for the Fornax periodic
variables compared to the PL for C-rich Miras in the LMC,
$M_K=-3.51[\log P-2.38]-7.24$, from Whitelock et al. (2008)
on the assumption that the Fornax distance modulus is 20.69 mag (see section
8). Four of the Miras fall well below the LMC PL relation.
Fig.~\ref{fig_hkp} shows that the four faint Miras are the reddest and, with
the exception of the longest period object, are significantly redder than
their LMC counterparts which were used to establish the $K$ PL relation. 
Their colours suggest that circumstellar extinction will affect the $K$
luminosity, to an extent that the $K$ PL relation will not be useful.

\begin{figure}
\includegraphics[width=8.5cm]{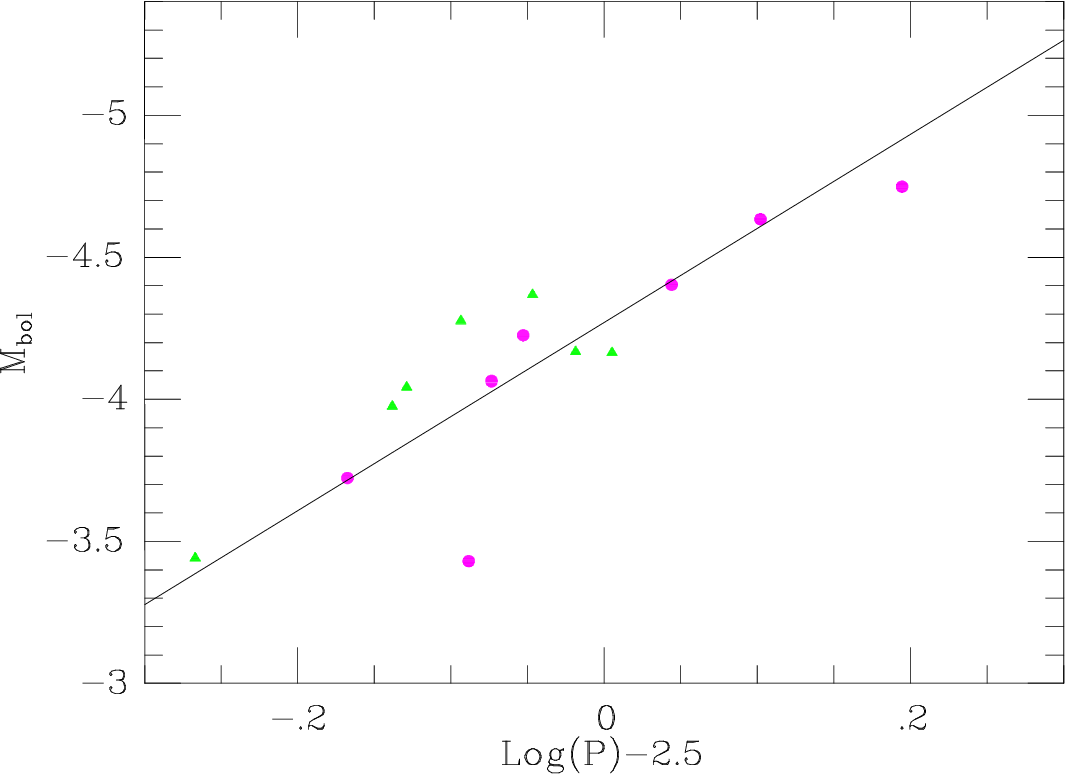}
\caption{The bolometric PL relation for the periodic variables in Fornax.
Miras are shown as circles and SRs as triangles. The line is the result of
fitting equation 1 to the 6 brightest Miras, which gives a distance modulus of
20.69 mag.}
\label{fig_pl}
\end{figure}

 Apparent bolometric magnitudes were calculated using the mean of the
bolometric corrections as a function of $(J-K)$ and $(H-K)$ (Whitelock et
al. 2006 equation 10 and table 5); the results are shown in
Table~\ref{tab_LPV}, column 9. 
Various observations, see, e.g. Whitelock et
al. (2008), suggest that the same PL relation applies to Miras in very different
environments. We therefore use the PL relation for the LMC (equation 1 of
the appendix) to find the distance modulus for Fornax. Using the 7 Mira
variables, we find $(m-M)_0=20.77\pm0.09$ mag (internal error). Omitting
F11020, which is clearly fainter than the others, we obtain
$(m-M)_0=20.69\pm0.04$ mag (internal error). Fig.~\ref{fig_pl} shows the PL
relation, using the distance modulus just derived from the six Miras. The
straight line is equation 1 from the appendix. Note that if the Kato et al. 
(2007) colour transformations were applied (see section 2) then it would 
make the distance modulus larger by only 0.01 mag.

For the absolute error we must take into account the scatter on the PL
relation determined in the appendix, $\pm 0.12$, and the uncertainty of the
adopted distance to the LMC, $18.39 \pm 0.05$ (van Leeuwen et al. 2007). 
Taking these into account the distance to Fornax is $(m-M)_0=20.69\pm0.08$
mag (or $(m-M)_0=20.77\pm 0.11$ mag, if the faint Mira is included). This is
in good agreement with other determinations, discussed below.

It is not certain why F11020 is fainter than the PL relation, but in view
of its colour it seems possible that it was experiencing an obscuration
event of the kind that is common in C-rich Galactic Miras (see also
section 7.1). Whitelock et al. (2006) estimated that at least one third 
of C-rich Miras in the Galaxy underwent such obscuration events, so
it is not surprising that one out of seven Miras in Fornax was caught in
the act of ejecting a puff into the line of site.

Matsuura et al. (2007) fit model SEDs to 2MASS data and mid-infrared Spitzer
spectra to estimate the bolometric magnitude of several stars in Fornax. The
three stars with the strongest mid-infrared emission are the three longest
period variables: F13023 (their 13-23), F12010 (their 12-4) and F3099 (their
3-129) for which they determine bolometric magnitudes of, --4.87, --5.49
and --4.92, respectively (after correcting for the differences in our assumed
distance modulus) that are significantly brighter than our values. Lagadec et
al. (2008)\footnote{Lagadec et al. have the two stars 13023 (For1) and 12010
(For2) confused in their table 2.} and Groenewegen et al. (2008) found
bolometric magnitudes roughly midway between the values from Matsuura et al. and
ours of $M_{bol}=-4.81$ and $-4.69$ (for F13023) and $-5.23$ and $-5.17$ (for
F12010), respectively, from the 2MASS magnitudes and the same $(J-K)$
dependent bolometric correction that we used, so about half the difference
between our values and those of Matsuura et al. can be attributed to the
bright 2MASS magnitudes. In fact the best agreement is for F3099 where the
2MASS magnitudes are close to our mean values, despite the fact that in this
star the mid-infrared provides a significant contribution to the total
luminosity. Note that a star with $M_{bol}=-5.49$ (the value Matsuura et al. 
found for F12010), would have a period of 740 days if it fell on a
period-luminosity relation. There is nothing in the $9\, \mu$m measurement
from Lagadec et al. to suggest extraordinarily strong mid-infrared emission,
so we suspect that using the code DUSTY to fit the SED for single epoch
observations obtained on widely different dates has led to an
unreasonably high total flux for these stars.

Once it is possible to monitor these extreme AGB stars around their light
cycles at mid-infrared wavelength we will be able to determine better mean
bolometric luminosities for them. That is, however, not critically important
for determining distances. For that purpose what is important is to follow
the identical procedure for the target stars, in this case the Fornax
Miras, and the calibrating stars, in this case the LMC Miras. If that is
done, as it has been here, differences in the SED of the target and
calibrating stars may be a small problem, but the major source of
uncertainty is the distance to the LMC itself.

Fig.~\ref{fig_pl} also shows the positions of the SR variables on the PL
relation and shows that some fall very close the relation defined by the
Miras. This is perhaps not surprising given the similarity of C-rich Miras
and SRs mentioned above. Those that fall above the PL are probably less
evolved and may be pulsating in the first overtone.
This can be compared to the situation for LMC variables where SR variables
are known to fall on various PL($K$) relations, mostly above that occupied by
Miras (Wood 2000; Ita et al. 2004 fig.~1).

\section{The Distance to Fornax}
 The literature contains a large number of estimates of the distance to the
Fornax dwarf spheroidal. It is not always straightforward to compare these
as they are often based on different or even mutually contradictory
assumptions. We therefore reanalyze the major works here and, as far as is
possible, put them on the same zero-point so that they may be inter-compared.

Table~\ref{distance} lists various recent distance estimates made by
different methods which can be categorized as follows: (1) the luminosity of
the TRGB; (2) the luminosity of the horizontal branch (HB) (including red
HB, the RR Lyrae variables (RRs) and RRs in clusters; (3) red clump
luminosity; (4)
$\delta$ Sct variables. The waveband in which the measurement is made is
given in brackets after the abbreviation of the method, e.g. $V$, $I$ or $K$.

As discussed in section 3 we find that the TRGB varies from $K_S\sim 14.5$
at $J-K_S=1.07$ to $K_S\sim 14.9$ mag at $J-K_S=0.81$ mag. These values
obviously bracket the $K=14.61$ mag adopted by Gullieuszik et al. 2007, who
quote errors of $\pm 0.02$ (random) and $\pm 0.03$ (systematic) associated
with their value.  It would obviously be possible to use our values to
determine a distance as did Gullieuszik et al. using an expression from
Valenti et al. (2004) that links the TRGB with the metallicity and apply a
correction for age. However, without a better indication and what the ranges
of age and metallicity actually are we don't believe that such an estimate
will significantly advance our understanding of the distance to Fornax.

The values are quoted in the column labeled ``$(m-M)$ orig", as they were
listed in the original paper, together with the error given. Many of the
standard errors are internal estimates and do not include the uncertainty of
the absolute calibration. For the TRGB, Bellazzini (2008) writes ``...the
Zero Point is known with uncertainties of $\pm 0.12$, in the best case; TRGB
distance moduli with error bars smaller than this figure neglect part of the
actual error budget". In the case of the $V$ observations, uncertainty in
the reddening is important (e.g. Rizzi et al. (2007b) note that their HB
modulus is reduced by 0.1 mag if $E(B-V)$ is 0.05 rather than 0.03).

 The TRGB moduli (items 1 to 3 in Table~\ref{distance}) and the HB moduli
(items 4 to 8 of Table~\ref{distance}) were calibrated by the original
workers using some adopted absolute magnitude for the HB or RR Lyrae
variables (see notes to Table~\ref{distance}). In the column ``$(m-M)$ rev"
these have all been reduced to a common zero-point based on the apparent
magnitudes of RR Lyrae variables in the LMC (Gratton et al. 2003) together
with an LMC modulus of 18.39 mag derived by van Leeuwen et al. (2007) using
parallaxes of classical Cepheids (see also Feast et al. 2008). Since these
estimates are not independent, they are average as indicated in the last
column of Table~\ref{distance} and this mean is given unit weight. The
results of the present paper depend on the same LMC modulus together with
data on the carbon-rich Miras in the LMC (Whitelock et al. 2008 and section
7.3 above). They are therefore partially independent of items 1 to 8 in
zero-point and are treated separately. The results for the red clump and for
$\delta$ Scuti stars depend on parallaxes of local stars of these types (see
notes to Table~\ref{distance}) and are taken directly from the original
sources. Each of the various methods in Table~\ref{distance} has its own
assumptions and uncertainties. Our final result depends on the mean of the
four estimates in heavy type in the last column of table 4. This is 20.69
mag with a range of 20.58 to 20.78. If the TRGB and HB were each given unit
weight the mean result would be 20.67 mag. Given the various uncertainties this
is not significantly different from our adopted value. Further improvement
in the Fornax modulus will come both from improvements in basic calibrations
and in greater understanding of the corrections necessary in at least some
of the cases for age and metallicity effects.

\begin{table*}
\begin{center}
\caption{The distance modulus for the Fornax dSph.}
\label{distance}
\begin{tabular}{llllllllll}
\hline
No. &
\multicolumn{1}{l}{Reference} & \multicolumn{1}{l}{Method} &
\multicolumn{1}{l}{$(m-M)$ orig} & \multicolumn{1}{l}{$(m-M)$ rev} &
\multicolumn{1}{l}{Means} \\
\hline
  1 & Gullieuszik et al. (2007) & TRGB($K$) &  $20.75\pm 0.19$ & 20.59\\
  2 & Rizzi et al. (2007a)      & TRGB($I$) &  $20.76\pm 0.04$ &  20.55\\
  3 & Rizzi et al. (2007b)      & TRGB($I$) &  $20.71\pm 0.07$ &  20.57\\
\multicolumn{5}{l}{Mean methods using the TRGB (1, 2 and 3) } & 20.57\\
  4 & Savianne et al. (2000) &  red HB($V$) &  $20.76\pm 0.04$ & 20.66\\
  5 & Greco et al. (2005) &   RRs($V$) &     $20.72\pm 0.10$ & 20.54\\
  6 & Rizzi et al. (2007b) &   HB($V$)  &     $20.72\pm 0.06$ & 20.6\\
  7 & Greco et al. (2007) &   cluster RRs($V$) & $20.64\pm 0.09$ & 20.55\\
  8 & Mackey \& Gilmore (2003) &   cluster RRs($V$) & $20.66\pm 0.15$ & 20.56\\
\multicolumn{5}{l}{Mean of methods using the HB (4, 5, 6, 7, 8)} & 20.59\\
\multicolumn{5}{l}{\bf Mean of methods using the HB and TRGB} & {\bf 20.58}\\
  9 & Pietrzynski et al. (2003) &  RC($K$) &  $20.86\pm 0.01$ & 20.86\\
 10 & Gullieuszik et al. (2007) & RC($K$)  &  $20.74\pm 0.11$ &  20.74\\
 11 & Rizzi et al. (2007b)   &  RC($I$)    &  $20.73$     &     20.73\\
\multicolumn{5}{l}{\bf Mean of methods using the RC (9, 10, 11)} & {\bf 20.78}\\
 12 & Poretti et al. (2008) & $\delta$ Sct  &  $20.70\pm 0.02$ & 20.70\\
\multicolumn{5}{l}{\bf Method using $\delta$ Sct stars (12)} & {\bf 20.70}\\
13 & This paper & Mira PL & $20.69\pm0.04$ & $20.69\pm0.08$\\
\multicolumn{5}{l}{\bf Method using Mira PL (13)} & {\bf 20.69}\\
\multicolumn{5}{l}{\bf Mean of 4 different methods:} & {\bf 20.69}\\
\hline
\end{tabular}
\end{center}
\flushleft
Notes to Table~\ref{distance}:\\
{\bf  TRGB distances}: There are possible age and metallicity effects in these, 
in addition to the question of the absolute calibration. \\
  Mod 1. This is based on a theoretical ZAHB calibration. They estimate
 from the inferred chemical evolution history of Fornax plus models that
 the TRGB is 0.22 mag brighter there than for an old population, and
 apply this correction. \\
 Mod 2. This distance is based on a HB from  Carretta et al. (2000) (which contains
 a metallicity term) together with a metallicity dependent TRGB correction.
 They use [Fe/H] = --1.5. \\
 Mod 3. They use a ZAHB from Ferraro et al. (1999). From $V-I$ they estimate
        [Fe/H] = --1.5. This leads to $(m-M)_0 = 20.78$. But they finally
 adopt [Fe/H] = --1.0 and 20.71. \\
{\bf Horizontal Branch distances} \\
 Mod 4. Assumes [Fe/H] = --1.8 and a RR Lyrae/HB relation. \\
 Mod 5. They adopt an RR relation and [Fe/H] = --1.78 and  $E(B-V) =0.04$. \\
 Mod 6. They use an HB relation. \\
 Mod 7. Note especially their table 5 which lists Mod from HB/RR for
 various Fornax fields and globular clusters. This shows the range of
reddenings used. These authors use E(B-V)= 0.10 and 
adopt [Fe/H] = --2.01 for their cluster. \\
 Mod 8. These authors find distance moduli for four clusters ranging from 20.59 to 20.75
 which they take as indicating a significant depth, of 8 to 10 kpc, for Fornax. \\
{\bf Red Clump distances}\\
 Mod 9. Calibration is from Alves (2000) using Galactic parallaxes. These
authors find there is little
or no population correction necessary for this indicator. In principle
this is independent of our distance calibration etc. Note they find an
0.2 mag difference in $K$ and $I$ clump distances locally from that in
LMC, SMC, Fornax and Carina which the authors attribute to photometric errors
in the Galactic stars.\\
 Mod 10. This adopts the same calibration as Mod 9 except that they apply a population 
 correction based on models.\\
 Mod 11. This also applies a population correction based on models.\\
{\bf $\delta$ Sct star distances}\\
 Mod 12.   This depends on a PL[Fe/H] relation and adopting [Fe/H] = --1.4.
       The relation is based on Galactic $\delta$ Sct stars with 
 Hipparcos parallaxes, though no details of this have yet been published 
(see McNamara et al. 2007). 
 Poretti et al. point out some uncertainties with this calibration,  
 especially the problem of mode identification in the
       low amplitude Galactic stars (e.g. their section 6.2). The
method probably also needs re-examining using the revised 
Hipparcos results (van Leeuwen 2007).
 \end{table*}


\section*{Acknowledgments} We are grateful to the following people for
acquiring images for this programme: Enrico Olivier, Hirofumi Hatano,
Shogo Nishiyama, Hideyuki Shimizu. We also thank Martin Groenewegen
and colleagues for sending us their paper in advance of publication,
Hinako Fukushi for providing information on sources detected 
by AKARI and the referee, Jacco van Loon for some helpful suggestions.
 
This research has made use of Aladin. 
This publication makes use of data products from the Two Micron All Sky
Survey, which is a joint project of the University of Massachusetts and the
Infrared Processing and Analysis Center/California Institute of Technology,
funded by the National Aeronautics and Space Administration and the National
Science Foundation.

\section{Appendix: LMC PL}

In this paper we make use of bolometric magnitudes derived by applying a
$J-K$ and a $H-K$ dependent bolometric correction (Whitelock et al. 2006 equation 10) to
the mean $K$ mag. In order to use these with a PL relation we re-derive the
LMC PL relation using bolometric magnitudes derived in the same way.

For this we use only C-rich Mira variables taking the data from Feast et al.
(1989), revised according to the periods and classifications given by
Glass \& Lloyd Evans (2003)\footnote{for GRV0519--6454 the period from Glass et
al. (1990) is used.} together with the 5 C-Miras from Whitelock et
al. (2003) which have sufficiently good $J$ mags to derive the bolometric
correction. The Feast et al. data were converted onto the SAAO system
(Carter 1990). Interstellar extinction was corrected for as described by
Feast et al. (1989).

If we take $(m-M)_0= 18.39$ mag for the LMC, then
fitting a PL relation to these 22 points provides the following:
\begin{equation}
M_{bol}= -4.271(\pm 0.026)-3.31(\pm 0.24)[\log P -2.5],
\end{equation}
neglecting the error on the distance modulus of the LMC. The rms scatter
about this relation is 0.12 mag. Compared with equation 1 of Whitelock et
al. (2006), after correcting to the same LMC distance, this new relation
will provide bolometric magnitudes that are 0.18 mag fainter at $\log P
=2.4$ and 0.05 mag brighter at $\log P=2.7$.

\end{document}